\documentclass[a4paper,fleqn,usenatbib]{mnras}
\usepackage[T1]{fontenc}
\usepackage{ae,aecompl}
\usepackage{natbib}
\bibpunct{(}{)}{;}{a}{,}{,}

\usepackage{lmodern}
\usepackage[T1]{fontenc}

\usepackage{graphicx}	% Including figure files
\usepackage{amsmath}	% Advanced maths commands
\usepackage{amssymb}	% Extra maths symbols
\usepackage{multirow}
\title[Nucleosynthesis in advective disc and outflow]{Nucleosynthesis in advective disc and outflow: possible explanation for overabundances in winds from X-ray binaries}

\author[Datta \& Mukhopadhyay]{
Sudeb Ranjan Datta,\thanks{E-mail: sudebd@iisc.ac.in (SRD)}
Banibrata Mukhopadhyay\thanks{E-mail: bm@iisc.ac.in (BM)}
\\
Department of Physics, Indian Institute of Science, Bangalore 560012, India
}

\date{Accepted XXX. Received YYY; in original form ZZZ}

\pubyear{2015}

\begin{document}
\label{firstpage}
\pagerange{\pageref{firstpage}--\pageref{lastpage}}
\maketitle

\begin{abstract}
Multiple spectroscopic lines of different elements observed in winds from X-ray binaries (XRBs), based on one zone model, indicate super-solar abundance of elements, e.g. Mg, Si, S, Ar, Ca, Cr, Mn, Co. The one zone model considers similar hydrodynamics of underlying winds. In order to find a possible origin of these overabundances, we explore nucleosynthesis in advective, geometrically thick, sub-Keplerian, accretion disc in XRBs and active galactic nuclei (AGNs), and further in outflows launched from the disc. Based on flow hydrodynamics and solving
nuclear network code therein by semi-implicit Euler method, we obtain abundance
evolution of the elements. Although the density is very low, due to very high temperature of advective disc than Keplerian Shakura-Sunyaev disc (SSD), it is quite
evident that significant nucleosynthesis occurs in the former. As the temperature at
the base of the outflow is constrained by the temperature of disc, nucleosynthesis also occurs in the outflow contingent upon its launching temperature.
Till now, the outer region of XRB and AGN discs is understood to be 
colder SSD and inner region to be advective disc, together forming a disc-wind system.
Hence, newly evolved abundances after processing through outflow can change the
abundances of different elements present in the environment of the whole disc-wind
system. We find 2-6 times overabundant Mg, Si, Ar, Cr with respect to the respective solar abundances, which is consistent observationally. Thus for most XRBs, when only iron lines are
present, inclusion of these evolved abundances is expected to change the observational
analysis drastically.
%This is a simple template for authors to write new MNRAS papers.
%The abstract should briefly describe the aims, methods, and main results of %the paper.
%It should be a single paragraph not more than 250 words (200 words for %Letters).
%No references should appear in the abstract.
\end{abstract}


%%%%%%%%%%%%%%%%%%%%%%%%%%%%%%%%%%%%%%%%%%%%%%%%%%

%%%%%%%%%%%%%%%%% BODY OF PAPER %%%%%%%%%%%%%%%%%%

%\section{Introduction}

%This is a simple template for authors to write new MNRAS papers.
%See \texttt{mnras\_sample.tex} for a more complex example, and \texttt{mnras%\_guide.tex}
%for a full user guide.

%All papers should start with an Introduction section, which sets the work
%in context, cites relevant earlier studies in the field by \citet{Others2013},
%and describes the problem the authors aim to solve \citep[e.g.][]{Author2012}.

\section{Introduction}
The advective accretion disc arises when accretion rate is low. Due to their low density and shorter residence time, protons are unable to transfer their energy to electrons and two-temperature plasma comes in the picture. Such discs are optically thin, geometrically thick and their ion temperature could reach the virial value $ \sim (1/k)(GMm_p/r_s) \sim 5.4 \times 10^{12}$ K, if there is no cooling. Here $k$ and $G$ are the Boltzmann and gravitational constants, $M$ is the mass of black hole, $m_p$ is the mass of proton, $r_s=2GM/c^2$ is the Schwarzschild radius and $c$ is the speed of light in vacuum. Realistically, electron temperature goes to $ \sim 10^9$ K, whereas ion temperature reaches to $ \sim 10^{11}$ K. This temperature is too high for nucleosynthesis, i.e. at this temperature all the nuclei will be disintegrated into free nucleons$-$neutrons and protons.

On the other hand, there is another accretion disc regime, which is most established, known as Shakura-Sunyaev disc (SSD) model (\citealt{1973A&A....24..337S}). SSD is optically thick and geometrically thin and arises when accretion rate is relatively high. Due to large optical depth, it radiates as multi-temperature blackbody and temperature attains at most $ \sim10^7$ K. Although in this disc, density is higher, temperature is too low for nucleosynthesis, because reaction rates are highly sensitive to the temperature.

In between these two extreme regimes, there is an accretion zone which is also sub-Keplerian and advective in nature. The temperature therein is to be adequate to kick in nucleosynthesis. In very inner region of such regime, the temperature may turn out to be too high to disintegrate all the nuclei, as mentioned above.

Till now the best understanding is that in the outer region of the X-ray binary accretion disc around a compact object (XRB), the colder Keplerian SSD is present and in the inner region, sub-Keplerian hotter advective disc is present (e.g. \citealt{1995ApJ...455..623C,1998tbha.conf..148N,2013FrPhy...8..630Z}, and references therein). The same is true for accretion disc around a supermassive black hole, i.e. in an active galactic nucleus (AGN). We see XRBs in soft or hard states, depending on the dominance of SSD or advective disc components respectively. Therefore, considering only the temperature, it is justified to say that in accretion disc, there will be nucleosynthesis, i.e. abundance of the flow will evolve, as it goes from outer edge to the near black hole.

Many earlier works were devoted to nucleosynthesis in a XRBs as well as AGNs. \cite {1987ApJ...313..674C} and \cite {1989ApJ...336..572J} basically initiated nucleosynthesis in an optically thick SSD. They found that significant nucleosynthesis happens in the disc only when the viscosity parameter $\alpha_{visc}\leqslant10^{-4}$. This is because temperature and density increase and radial velocity decreases with decreasing $\alpha_{visc}$. However, this range of $\alpha_{visc}$ is not consistent observationally (\citealt{2007MNRAS.376.1740K}). \cite {1999A&A...344..105C} studied nucleosynthesis in hot, highly viscous advective accretion flows with small accretion rates and showed that neutron tori can form around a black hole due to disintegration of all the elements into free nucleons, which matches with the results presented in this work. \cite {2000A&A...353.1029M} studied nucleosynthesis in advective discs for other parameter space with and without the formation of shock in accretion discs. They found that significant nucleosynthesis can occur in the disc and in principle this can affect the metallicity of the galaxy. If we observe a specific XRB, then this change may become very much significant and analysis of observational data, ignoring these evolved abundances, can lead to very different results than the actual scenario. This is because, metallicity is a key factor to determine density and other physical parameters of the flow from the spectra.\\

GX 13+1 and GRO J1655-40 are the two Low Mass X-ray Binary (LMXB) sources, which exhibit the presence of many spectroscopic lines in the wind from accretion discs (\citealt {2004ApJ...609..325U}, \citealt {2012A&A...543A..50D}, \citealt {2009ApJ...701..865K}, \citealt {2018ApJ...861...26A}). To explain the strength of spectroscopic lines based on one absorption zone model, overabundance of some elements is required in comparison with the respective solar abundances. It is very unlikely that these overabundances are caused by either initial abundance of the companion (\citealt {2015A&A...582A..81J}) or nucleosynthesis in supernova which is occurred at the time of formation of the compact objects (\citealt {2009ApJ...701..865K}). Hence, the question remains: from where these overabundances arise. We propose that these overabundances occur due to nucleosynthesis in an advective disc and associated outflow, which further leads to the enrichment of the environment of outer Keplerian part of the disc. The observationally inferred overabundances of Mg, Si, Ar and Cr match with those produced in our simulation quite well. However, the situation is really complex and theoretical calculation does not explain the overabundances of many elements like Ca, Ar, Mn and Co, which are also observed. Above facts also argue that nucleosynthesis has a potential to change the observational analysis drastically, when only Iron lines are present in the spectra, which is actually the case for most of the sources. Because, for those cases, to find out density and other physical variables of the wind as well as disc, we need to know how much fraction Iron is of total. Also we see in our simulation that in the inner region of the disc, all the elements are disintegrated into free nucleons. This can be a possible reason for spectroscopic lines to be indiscernible when the disc is in a pure hard state.\\

Nucleosynthesis in the AGN disc is also attempted to explore after developing the XRB cases. However, quantitative comparison of abundances in our simulation with observation for AGN is quite complex and involved (\citealt {1999ARA&A..37..487H}).\\

However, all the earlier explorations of nucleosynthesis in XRBs and AGNs are based on nucleosynthesis in the disc only (\citealt {1992A&A...254..191A, 2000A&A...353.1029M, 2018MNRAS.481L.110B}). The nucleosynthesis in the outflow was not considered self-consistently, when, however, flow cools down as it expands. The density and temperature at the base of outflow are constrained by the density and temperature of the disc. Therefore, whenever there is an abundance evolution in the disc, there may be some evolution in the outflow, contingent upon density and temperature at the launching radius, which we consider here. For the present purpose, we use one of the simplistic spherically expanding adiabatic outflow models, which has been used extensively to study nucleosynthesis in the outflow from collapsar discs in past (\citealt {2004ApJ...614..847F, 2013ApJ...778....8B}). Although \cite {2008ApJ...681...96H} studied nucleosynthesis in the spherically expanding outflow from advection dominated accretion flows, their disc model is self-similar, which, we know, differs very much from the actual solution in the inner region of the disc. They also used the scaling relation for expansion time ($\tau$) with density ($\rho$) as $\tau\sim446/\rho^{1/2}$ for the dynamical evolution of outflow.\\

We consider here the magnetized advective accretion disc solution following \cite{2018MNRAS.476.2396M} (hereafter MM18) for the present purpose of nucleosynthesis in XRBs and AGNs. However, the results do not differ 
dramatically for any other self-consistent advective disc models. We follow the geometry of the outflow 
to find out the velocity of the expansion, which is more realistic with the outflow picture. Finally, 
our result reveals that final abundance in outflow depends mostly on the velocity of the outflow with 
which it launches, and not really on its density and temperature profiles. Numerical general
relativistic magnetohydrodynamic (GRMHD) simulations 
(\citealt{2007MNRAS.375..513M,2012MNRAS.426.3241N,2014MNRAS.439..503S}) indicate 
us to choose the appropriate outflow velocity.
As shown in the subsequent sections, if outflow velocity is small ($\sim 0.005 c$ or less), then the final 
abundance changes drastically from the disc. However, if the velocity is sufficiently high, then the final 
abundance remains almost same as that at the launching radius of the disc. This argues that we should not 
conclude about the elements and elemental synthesis by which outer Keplerian part as well as surrounding 
environment enriched by performing nucleosynthesis only in the disc. The outflow can influence 
elemental synthesis and final abundance substantially.

The plan of the paper is as follows. In Section 2, we present how theoretically and computationally we perform nucleosynthesis. We describe models of disc and outflow in Section 3. In Section 4, we present the results of nucleosynthesis for XRB as well as AGN cases. We have described about the shortcomings of our model and compare our results with observation in Section 5. We justify the choice of simplistic outflow model for our work in Section 6. Finally we conclude in Section 7.

\section{Theoretical framework}

In this section, we discuss how theoretically and computationally we mimic the physical situation and how the nucleosynthesis is performed. First of all, we need initial abundances over which reactions to occur. Choosing initial abundance is a very crucial step as depending on this all the reactions occur. However, as all the elements finally are disintegrated to free nucleons in the inner region of the flow, final results will not depend critically on the initial abundances unless it is drastically different. Nevertheless, initial abundance determines where in the disc various reactions occur. It is useful to represent the abundances in terms of mass fraction $(X)$.

After choosing initial abundances for all isotopes, we need temperature ($T$), density ($\rho$) and velocity ($v$) of the flow, i.e. hydrodynamics of the flow. As the plan is nucleosynthesis in advective disc and in outflow launched from the disc, we need hydrodynamic profiles of advective disc and outflow which we discuss in the next section. Reaction rates are taken from JINA reaclib database\footnote{https://groups.nscl.msu.edu/jina/reaclib/db/}(\citealt {2010ApJS..189..240C}). For flexibility and to use in larger network, the rate of the $i$-th reaction is presented following \cite {1980PhDT.........1T} as

\begin{multline}
N_A<\sigma v>_i=exp[c^1_i+c^2_i T^{-1}_9+c^3_i T^{-1/3}_9+c^4_i T^{1/3}_9+c^5_i T_9+c^6_i T^{5/3}_9 \\
+c^7_i log(T_9)],
\label{Thielemann coefficients}
\end{multline}
where $N_A$ is the Avogadro number, $<\sigma v>_i$ is the reaction rate per unit volume, $c_i$-s are the Thielemann coefficients, $T_9$ is the temperature in units of $10^9$ K. The values of coefficients $c_i$ are given in the JINA reaclib database for various reactions. It is clear from equation (\ref{Thielemann coefficients}) that reaction rates are highly sensitive to the temperature.\\

Total number of reactions in an unit time step for a specific element depends on how many nuclei of that element present in the system. The requirement of density comes here. The mass fraction of each element depending on density gives the number of nuclei of the respective element which determines how much each element will be destroyed/synthesized in an unit time step.\\
 
As the flow is moving towards a black hole, we need to keep track of how much time it is spending in a specific temperature and density region. If the velocity is low, then elements in the flow will spend more and more time around a region, more and more reactions will occur in that region and abundance will evolve largely. Otherwise, if the velocity is very high, then elements in the flow will spend less and less time with insignificant number of reactions, which will keep the final abundance similar to the initial abundance. It has been found in our study that this effect is very crucial in determining the final abundance.\\
 
Now briefly we describe how reaction network changes the abundance following \cite {2006ApJ...642..443C}. For simplicity we assume 4 elements and 3 reactions: $^1H$($^1H$,$\beta^+\nu)D$($D$,$\gamma$)$^4He$($2^4He$,$\gamma$)$^{12}C$. Neglecting any backward reactions, the corresponding rate equations can be expressed as

\begin{equation}\label{reaction network matrix}
\frac{d}{dt}\begin{bmatrix}
X_H \\ X_D \\ X_{He} \\ X_C
\end{bmatrix}  =  \begin{bmatrix}
-\lambda_H & 0 & 0 & 0 \\ \lambda_H & -\lambda_D & 0 & 0 \\ 0 & \lambda_D & -\lambda_{He} & 0 \\ 0 & 0 & \lambda_{He} & 0
\end{bmatrix}
\begin{bmatrix}
X_H \\ X_D \\ X_{He} \\ X_C
\end{bmatrix}.
\end{equation}
Here $\lambda$-s are the reaction rates of specific reactions, subscripted by the respective main reactant. Therefore, if the temperature, density and velocity of the flow are available, then at each time step, solving the above set of equations using semi-implicit Euler method, we obtain the change of abundances at that step. Then updating the initial abundance for the next time step, which is same as the final abundance of the previous step, we obtain the evolution of elements through the whole flow. For the present purpose, we consider a reaction network of 3096 isotopes of different elements from Hydrogen to Copernicium for XRBs and a network of 366 isotopes from Hydrogen to Yttrium for AGNs. We have verified that our nuclear reaction network conserves mass to within one part in $10^8$. This is the same well tested nuclear network code which was implemented by \cite {2013RAA....13.1063B, 2013ApJ...778....8B} for studying nucleosynthesis related to the collapsar disc.\\

\section{Disc and outflow models}

Advective disc hydrodynamics is taken from MM18. The authors there discussed about the importance of large scale 
strong magnetic field in removal of angular momentum, as well as the possible origin of different kinds 
of magnetic barriers in advective, geometrically thick, sub-Keplerian accretion flows around black holes. 
We consider 3 different hydrodynamic profiles depending on different field configurations, as described
by MM18. The different sets considered are as follows.\linebreak\linebreak
 Set 1: $B_{rc}=2B_{\phi c}$; Set 2: $B_{rc}=0.5B_{\phi c}$; Set 3: $B_{rc}=B_{\phi c}$; \linebreak
 where $B_{rc}$ and $B_{\phi c}$ are the magnetic fields in the radial and azimuthal directions respectively 
at the critical point. It is important to emphasis that MM18 considered the strong field magnetic 
shear in order to remove angular momentum and subsequently accretion of matter. As the strong field is 
expected to sluggish magnetorotational instability (MRI), the effect of $\alpha_{visc}$ is expected 
to be sluggish too. Hence, MM18 did not consider $\alpha_{visc}$ in their model. On the other
hand, MM18 model is vertically averaged and hence it is unable to capture vertical motion/variation of the
flow (and flow variables), which is a drawback. Moreover, the vertical component of magnetic field was 
assumed to be zero. For more details, see MM18. Many of these shortcomings were removed by \cite{2019MNRAS.482L..24M}.

As shown by \cite{2015ApJ...807...43M} for weaker field magnetic shear, an efficient removal of 
angular momentum via magnetic shearing stress is possible and the flow is equivalent to that 
governed by MRI with $0.01 \lesssim\alpha_{visc}\lesssim 0.1$, depending on how the field varies vertically.  
From the comparison of flow profiles given by earlier works (e.g. \citealt{2000A&A...353.1029M,2010MNRAS.402..961R}),
we may anticipate the effective $\alpha_{visc}$ in MM18 to be $\gtrsim 0.05$.
However, for a more accurate determination of equivalent $\alpha_{visc}$ giving rise to similar
flow/hydrodynamic variables as of MM18, we need to solve disc equations, once considering shear only 
due to $\alpha_{visc}$, and then only with strong magnetic shear, and
adjust the $\alpha_{visc}$ in the former in order to produce flow profiles very
similar to the later, as was performed by \cite{2015ApJ...807...43M}.

\begin{figure*}

	\includegraphics[scale=0.6]{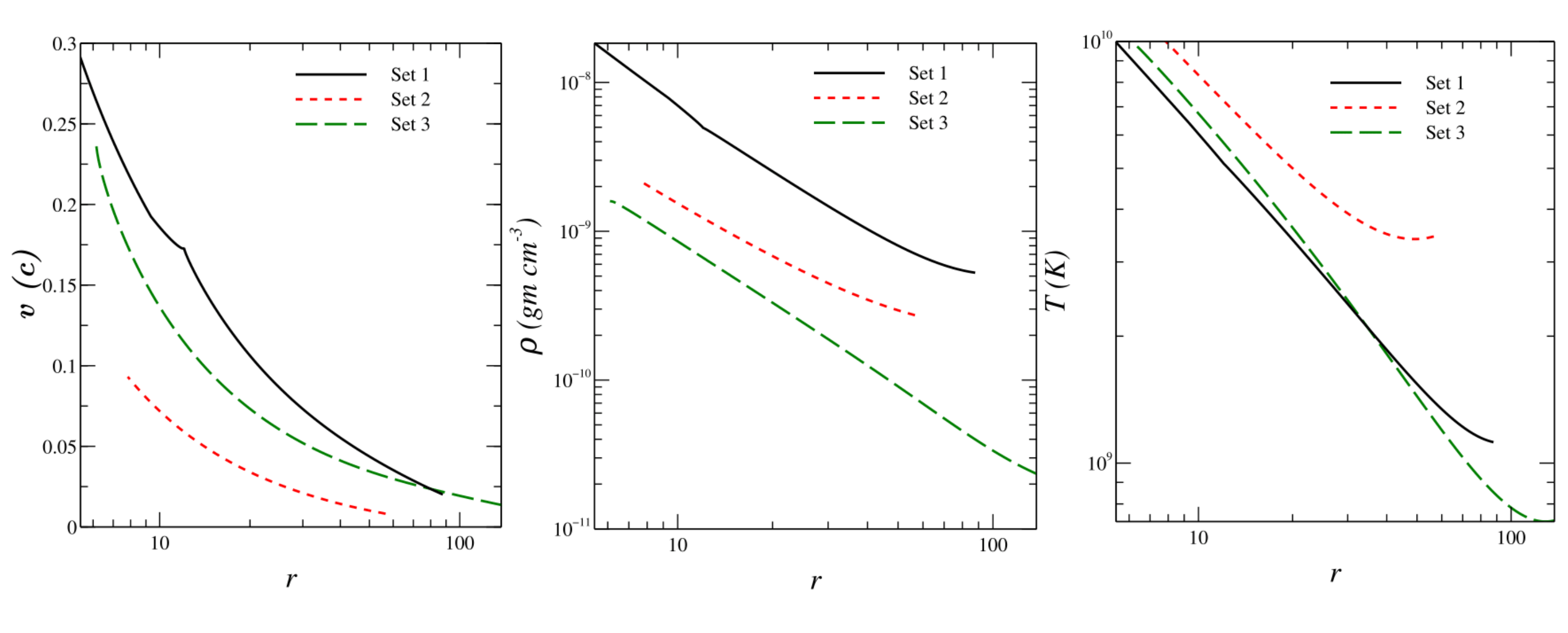}
    \caption{Velocity, density and temperature variations in advective accretion disc for three different sets.}

\label{Hydrodynamic profiles}
\end{figure*}
 
For the computation purpose, we use Schwarzschild radius $r_s (=2GM/c^2)$ as the unit of length scale. The mass of the black hole is assumed to be $10M_\odot$ for XRBs. The hydrodynamic solutions for advective disc are obtained by taking crudely the advection factor to 0.5, not by considering rigorous cooling effects. This factor includes all the information of cooling and hence advection. In the equation of state, MM18 used adiabatic index $\Gamma\sim$ 1.34, which corresponds to the ratio of gas to total pressure $\beta\sim$ 0.04. From the gas pressure, we determine temperature, which becomes $10^{9}-10^{10}$K. Fig. \ref{Hydrodynamic profiles} reveals the temperature, density and velocity profiles as functions of distance from the black hole ($r$) for the three above mentioned sets. It is quite natural that temperature, density and velocity increase as the flow advances from outer region to the near black hole. As the density is very low, the very high temperature becomes the key ingredient for the evolution of abundances of various elements.\\

For a supermassive black hole of mass $10^{7}M_\odot$ (i.e. AGN), the hydrodynamic variables in accretion discs with the same physical parameters as of XRBs remain same in the present model, except density scaling down with $M$. We therefore also explore, how only the change in mass of central black hole does affect the nucleosynthesis. Other than the change in density, another important change is the residence time of matter in the disc which scales as $M$. As we compare below, this makes the abundance evolution in AGNs different than XRBs.\\

For outflow, we assume simplistic spherically expanding adiabatic outflow model (\citealt {1996ApJ...471..331Q}). Although this model is used mainly for supernova discs (\citealt {2003ApJ...586.1254P}, \citealt {2013ApJ...778....8B}), for its simplistic picture, we use the model as the possible outflow from XRBs and AGNs. The adiabaticity fixes the temperature evolution as well as the entropy of the flow. The temperature of the ejecta based on this model is 
\begin{equation}\label{Temperature evolution}
T_{ej}(t)=T_0\left(\frac{R_{ej}}{R_{ej}+v_{ej}t}\right)^{3(\Gamma-1)},
\end{equation}
where $T_{ej}(t)$ and $T_0$ are the temperatures at time $t$ and $t$=0 respectively. Basically $T_0$ is the temperature of the disc where from outflow launches, $R_{ej}$ is the corresponding radius of the disc, $v_{ej}$ is the ejecta (outflow) velocity and $\Gamma$ is the adiabatic index of the flow. \\
 
The entropy of the ejecta is fixed at the value given by
\begin{equation}\label{Entropy}
\frac{S}{k}$$\approx$$0.052\frac{T_{MeV}^3}{\rho_{10}}+7.4+ln\frac{T_{MeV}^{3/2}}{\rho_{10}}=S_0,
\end{equation} 
where $T_{MeV}$ is the temperature in units of MeV and $\rho_{10}$ is the density in units of $10^{10}gm\ cm^{-3}$. The first term of equation \eqref{Entropy} is the contribution from relativistic particles, e.g. $\gamma$, $e^+$, $e^-$, and last two terms represent the contribution from nonrelativistic heavy particles. Depending on the temperature and density of the disc at a radius from where outflow is ejected, the entropy becomes fixed. As the volume expands, the temperature evolves adiabatically. The density is obtained from equation \eqref{Entropy} for given $S$ and $T$. Therefore, now we need to find out only that how the volume expands, i.e. the velocity, for that we follow \cite{2013ApJ...778....8B} given below. The spherical expansion corresponds to
\begin{equation}\label{spherical expansion rate}
\dot{M}_{ej}=4\pi r^2\rho v_{ej},
\end{equation}
where $r^2=R_{ej}^2+H_{ej}^2$, $H_{ej}$ is the scale height of the disc at $R_{ej}$. On the other hand, the mass accretion rate at $R_{ej}$ is 
\begin{equation}\label{accretion rate}
\dot{M}_{acc}=4\pi R_{ej} \rho H_{ej} v_R,
\end{equation}
where $v_R$ is the radial velocity at $R_{ej}$ of the disc. Dividing equation \eqref{spherical expansion rate} by equation \eqref{accretion rate} we obtain
\begin{equation}\label{ratio ejection accretion}
\frac{\dot{M}_{ej}}{\dot{M}_{acc}}=\frac{v_{ej}}{v_R}\left(\frac{R_{ej}}{H_{ej}}+\frac{H_{ej}}{R_{ej}}\right).
\end{equation}
From the disc hydrodynamics, we know $v_R$ and $H_{ej}$ at $R_{ej}$. Therefore fixing the ratio $\dot{M}_{ej}/\dot{M}_{acc}$, we obtain the ejecta velocity. This ratio acts as a free parameter in the model. Different values to this ratio give different ejecta velocities and we study nucleosynthesis for different ejecta velocities. We find that the final abundance processing through outflow depends critically on $v_{ej}$. A full justification of using this outflow model is discussed in Section 6.

\section{Nucleosynthesis}

  \subsection{X-ray binary}
  
     \subsubsection{Nucleosynthesis in disc}
  
  \begin{figure*}

	\includegraphics[scale=0.6]{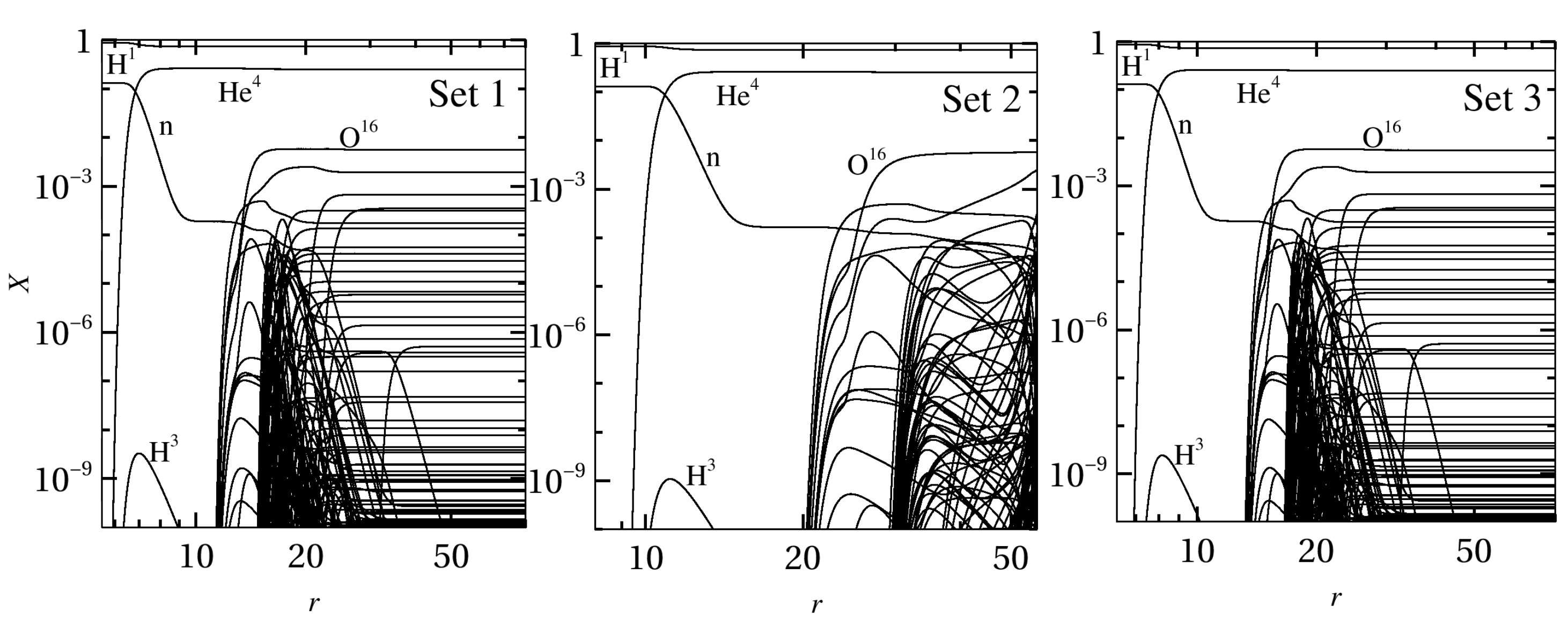}
    \caption{Abundance evolution of various elements in the disc for three different hydrodynamic sets. Some elements are explicitly denoted.}

\label{Disc chowmin picture}
\end{figure*}
  
The initial abundance of accretion disc for different isotopes is fixed as solar abundance, from \citealt {2009ARA&A..47..481A}. Some of the most abundant nuclei of Sun are tabulated in Table \ref{solar abundance}. Due to different temperature, density and velocity configurations of the disc flow, the initial abundances are evolved differently for different sets of disc profiles. However, the overall picture of disc nucleosynthesis remains same for all the sets. Initially, in the outer region of the advective disc, the temperature and density are so low for the Set 1 and Set 3 that no nuclear reaction takes place. As matter advances toward inner region of the disc, the temperature as well as density increase. Depending critically on the temperature, reaction rates become significant and abundances start to evolve. In the innermost region, all the elements finally are disintegrated to free nucleons. In the middle region of the flow, various reactions occur and abundances evolve in such a way that many peaks and dips of evolution of various elements occur (Fig. \ref{Disc chowmin picture}). The middle region comes in the picture because of the chain disintegration reactions, which eventually convert the initial elements to free nucleons. This scenario helps in capturing the right region of the accretion disc where nucleosynthesis occurs efficiently.\\

As disc hydrodynamic profiles for Set 1 and Set 3 are almost similar (Fig. \ref{Hydrodynamic profiles}), the abundance evolutions for these two sets (see Fig. \ref{Disc chowmin picture}) also crudely look similar. However, for Set 2, the size of advective disc is smaller and advective flow starts from a relatively higher temperature. Therefore, for Set 2, most of the heavy elements are disintegrated at the beginning of disc nucleosynthesis and very few elements remain in the disc with free nucleons. That is why Set 2 becomes less interesting to us. The whole picture of disc nucleosynthesis can be viewed typically as the disc nucleosynthesis of Set 1 or Set 3.\\

\begin{table}
	\centering
	\caption{Some most abundant nuclei present in solar abundance}
	\label{solar abundance}
	\begin{tabular}{lccr}
	\hline
	element & mass fraction & element & mass fraction\\
	\hline
	$H^1$ & 7.397$\times10^{-1}$ & $He^4$ & 2.509$\times10^{-1}$\\
	$C^{12}$ & 1.929$\times10^{-3}$ & $N^{14}$ & 6.723$\times10^{-4}$\\
	$O^{16}$ & 5.531$\times10^{-3}$ & $Ne^{20}$ & 3.473$\times10^{-4}$\\
	$Na^{23}$ & 2.958$\times10^{-5}$ & $Mg^{24}$ & 1.789$\times10^{-5}$\\
	$Al^{27}$ & 5.632$\times10^{-5}$ & $Si^{28}$ & 1.749$\times10^{-4}$\\
	$S^{32}$ & 1.359$\times10^{-4}$ & $Ca^{40}$ & 4.144$\times10^{-5}$\\
	$Mn^{55}$ & 1.095$\times10^{-5}$ & $Fe^{56}$ & 3.155$\times10^{-4}$\\
	\hline
	\end{tabular}

\end{table}

\begin{figure}

	\includegraphics[width=\columnwidth]{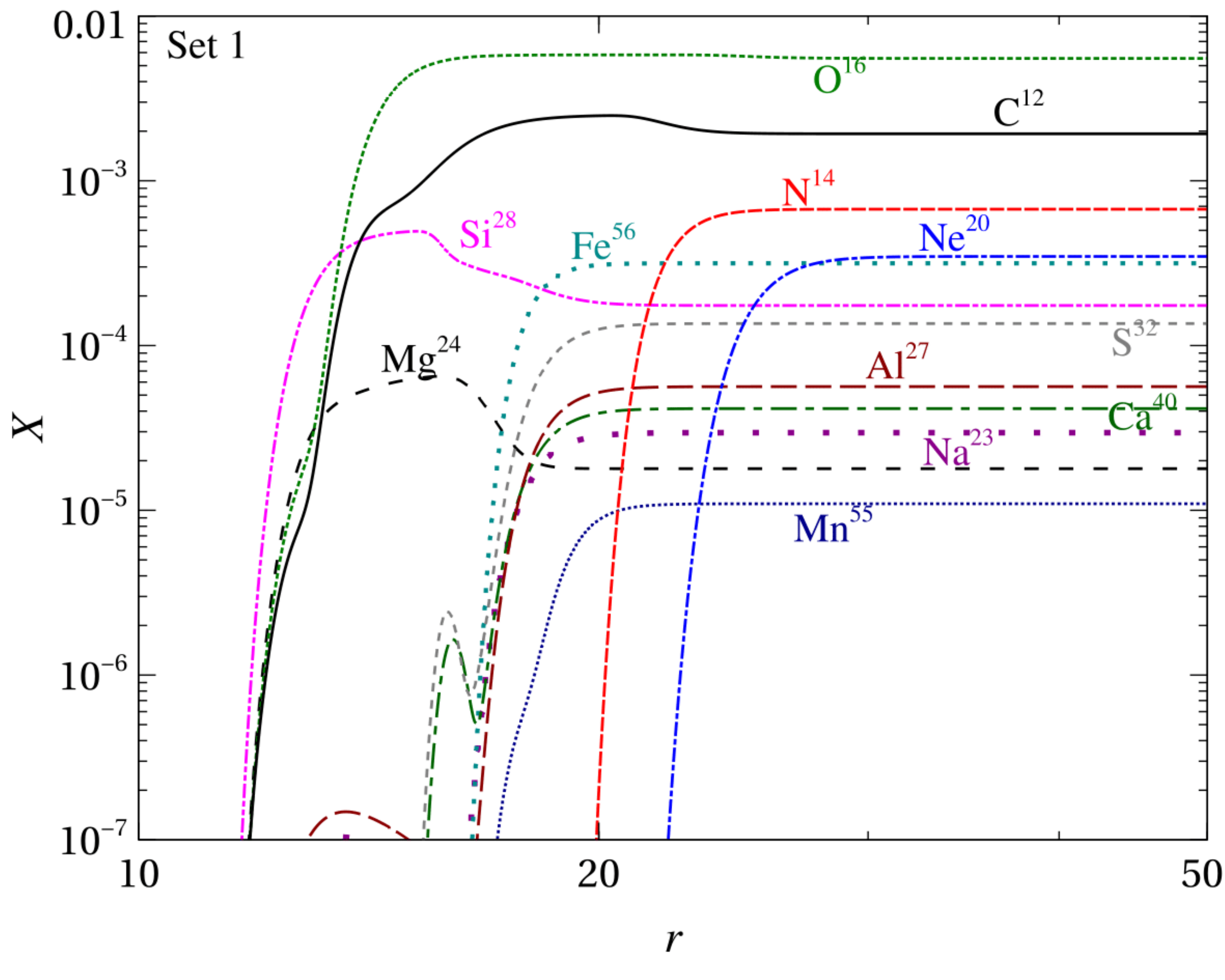}
    \caption{Disintegration of most abundant nuclei present initially (same as solar abundances) in XRB for Set 1.}

\label{disintegrated nuclei set1 disc}
\end{figure}

\begin{figure}

	\includegraphics[width=\columnwidth]{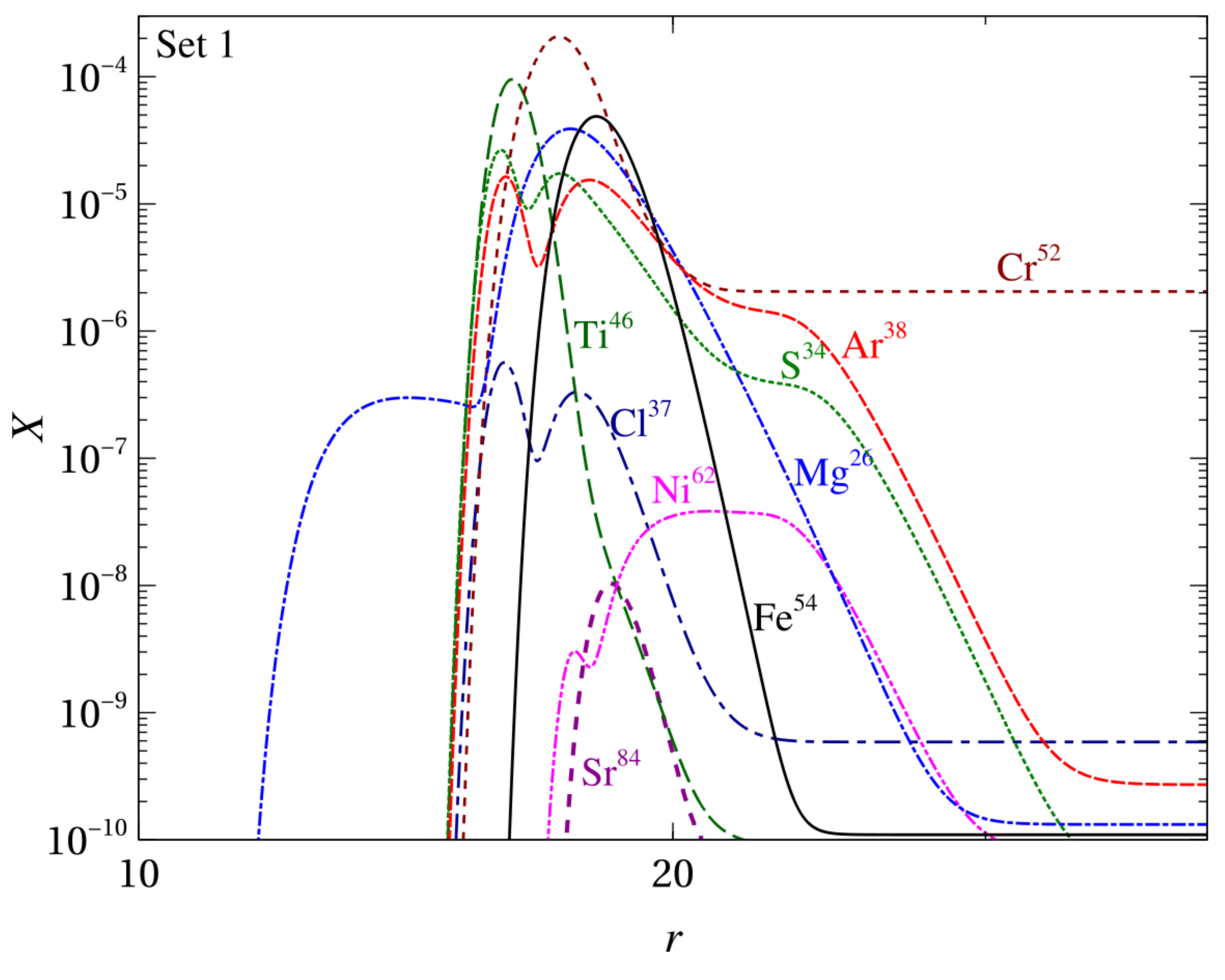}
    \caption{Formation and then disintegration of those elements which differ largely from initial solar abundance for Set 1 disc profiles in XRB.}

\label{peaks and dips of set1 disc}
\end{figure}

\begin{figure}

	\includegraphics[width=\columnwidth]{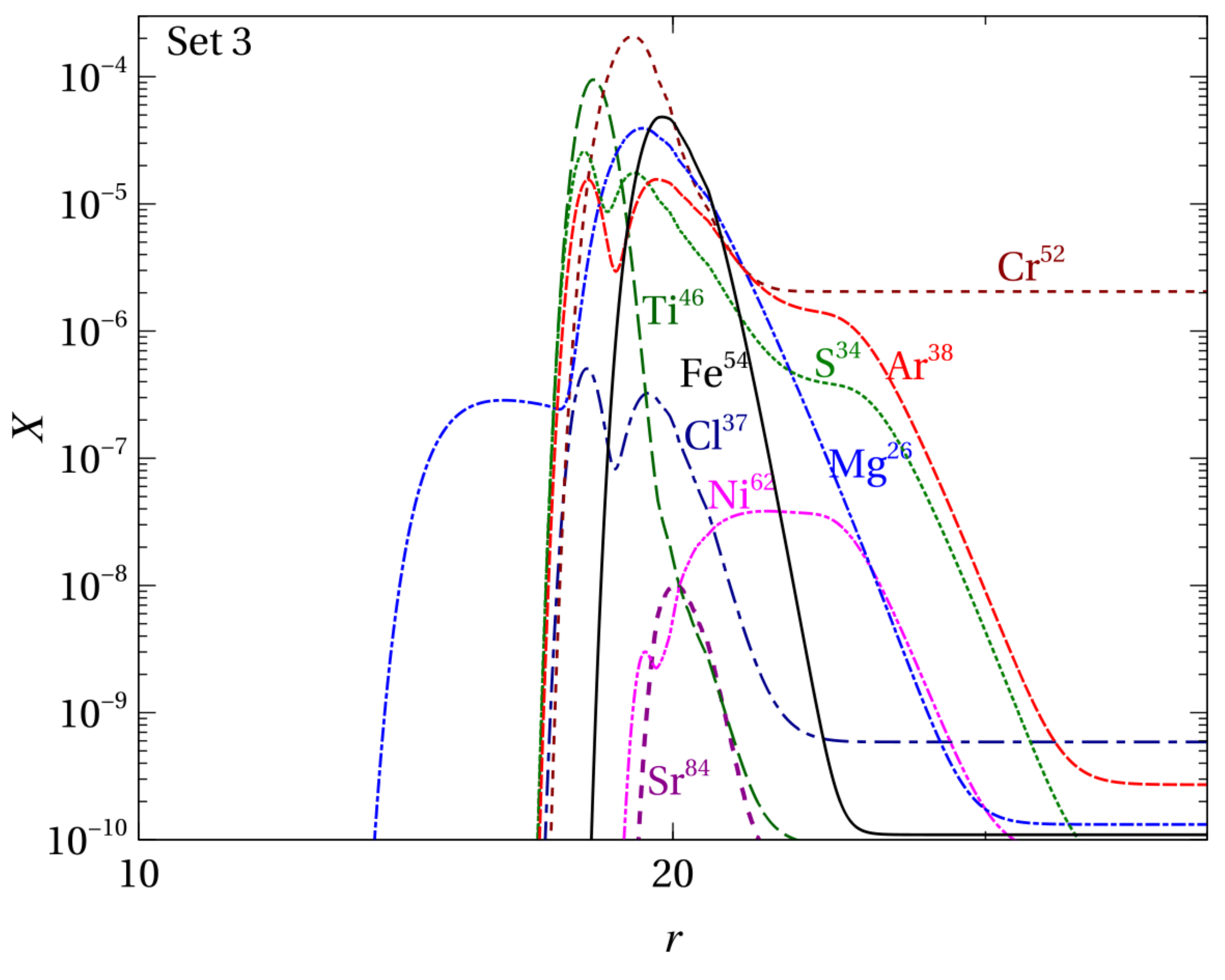}
    \caption{Same as Fig. \ref{peaks and dips of set1 disc}, but for Set 3.}

\label{peaks and dips of set3 disc}
\end{figure}

Some of the crucial chain disintegration reactions occurred in the disc are described as follows. The chain reaction in which $Fe^{56}$ disintegrates is: $Fe^{56}(n)Fe^{55}(n)Fe^{54}(H^1)Mn^{53}(H^1)Cr^{52}(H^1)V^{51}(H^1)Ti^{50}\\(n)Ti^{49}(n)Ti^{48}(n)Ti^{47}(n)Ti^{46}.$ $Mn^{55}$ disintegrates in the following way: $Mn^{55}(H^1)Cr^{54}(n)Cr^{53}(n)Cr^{52}$. Another chain reaction is $Ca^{40}(H^1)K^{39}(H^1)Ar^{38}(He^4)S^{34}(He^4)\\Si^{30}.$ $Ar^{38}$ disintegrates through another channel also given by, $Ar^{38}(n)Ar^{37}(n)Ar^{36}(H^1)Cl^{35}(H^1)S^{34}(He^{4})Si^{30}.$ Finally $Ne^{20}$ disintegrates to alpha particles through the reaction $Ne^{20}(He^4)O^{16}(He^4)C^{12}(2 He^4)He^4$. These reactions play the key role for the abundance evolution shown in Fig. \ref{Disc chowmin picture}. Disintegration of some of the most abundant nuclei in the disc for the profiles in Set 1 is shown in Fig. \ref{disintegrated nuclei set1 disc}. This picture remains almost same for Set 3. \\

Due to the above described chain reactions, various elements produce in the disc. The underlying evolutions are shown in Fig. \ref{peaks and dips of set1 disc} for Set 1 and in Fig. \ref{peaks and dips of set3 disc} for Set 3. All the elements are shown to differ largely from the initial abundances and cause the disc abundance to deviate from solar abundance. Note that the solar abundance is usually assumed to be granted for the analysis of observations of disc spectra. However, spectroscopic lines are observed, when sources are in between high/soft state (when Keplerian SSD is dominant) and low/hard state (when advective disc is dominant) (\citealt{2012MNRAS.422L..11P}), whereas we perform nucleosynthesis in the advective disc, which captures the required temperature suitable for nucleosynthesis.

The Bernoulli parameter for an advective disc generally remains positive (\citealt {1994ApJ...428L..13N}), which indicates that it is prone to outflow. This outflow is expected to be highly responsible to enrich the outer Keplerian disc by the newly generated nuclei. In the next section, we investigate the nucleosynthesis in outflow. Note importantly that advective disc remains dominant than Keplerian thin SSD in a low/hard state, which is expected to lead to non solar abundance in the disc in the low/hard state. One reason for not observing lines in a low/hard state could be due to the high temperature of the advective disc, when electrons can not remain in the atomic orbit for transition. The details of observational constraint and the presence/absence of winds in soft/hard states are discussed in the next section.\\

Figures \ref{peaks and dips of set1 disc} and \ref{peaks and dips of set3 disc} look almost similar, where the evolutions are for Set 1 and Set 3 respectively. The noticeable difference is, for Set 1 the evolution of isotopes leading to peaks spans in a larger range of radius than Set 3. This becomes clear from the hydrodynamic profiles for Set 1 and Set 3 (Fig. \ref{Hydrodynamic profiles}). In the radius range of evolution, the temperature is larger and the velocity is smaller for Set 3 relative to Set 1, which leads to maintain the flow at higher temperature for a longer time interval for the former. This is eventually responsible for the formation of those nuclei shown in Figs. \ref{peaks and dips of set1 disc} and \ref{peaks and dips of set3 disc} and then again due to higher temperature they disintegrate in the smaller range of radius for Set 3 compared to Set 1.\\

  \subsubsection{Nucleosynthesis in outflow}
  
When all the evolutions occur in the advective disc, some matter is ejected through the outflow. In this section, we discuss the nucleosynthesis in outflow. Finally outer Keplerian disc as well as nature will be enriched with those elements only, which come out from the advective disc. Initial abundances for nucleosynthesis in the outflow are the evolved abundances of the disc. Initial temperature and density of the outflow, when it is launched, are constrained by the temperature and density of the disc radius from where it is launched. Naturally, in outflow, the temperature and density decrease as it expands and goes away from the disc. \\

We have already divided the advective disc in three regions: outer region, middle region and inner region. When the outflow is launched from the outer region, we find that no nucleosynthesis takes place therein, because temperature in the disc itself is too low for nucleosynthesis, which decreases further in the outflow. When outflow is launched from the inner region where only free nucleons and helium nuclei survive, then also no significant nucleosynthesis takes place. Therefore, outflow from the inner region of the disc decreases the overall metallicity of the environment.\\

\begin{figure}

	\includegraphics[width=\columnwidth]{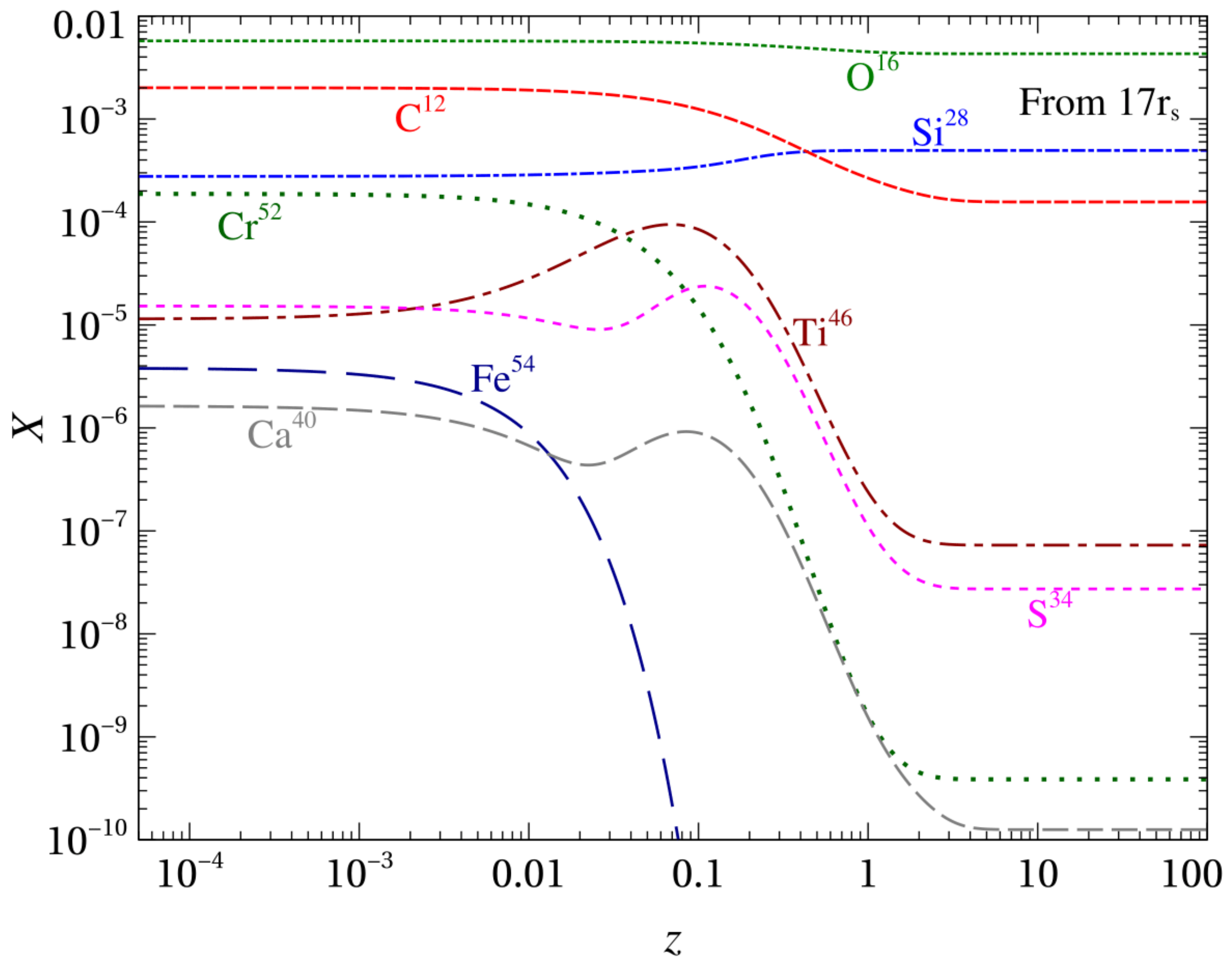}
    \caption{Evolution of abundances when outflow is launched from 17$r_s$ of Set 1 in XRB with $\dot{M}_{ej}/\dot{M}_{acc}=0.1$. Here $z$ is the distance from the accretion disc.}

\label{outflow set1 rea 0.1 17 rs}
\end{figure}

\begin{figure}

	\includegraphics[width=\columnwidth]{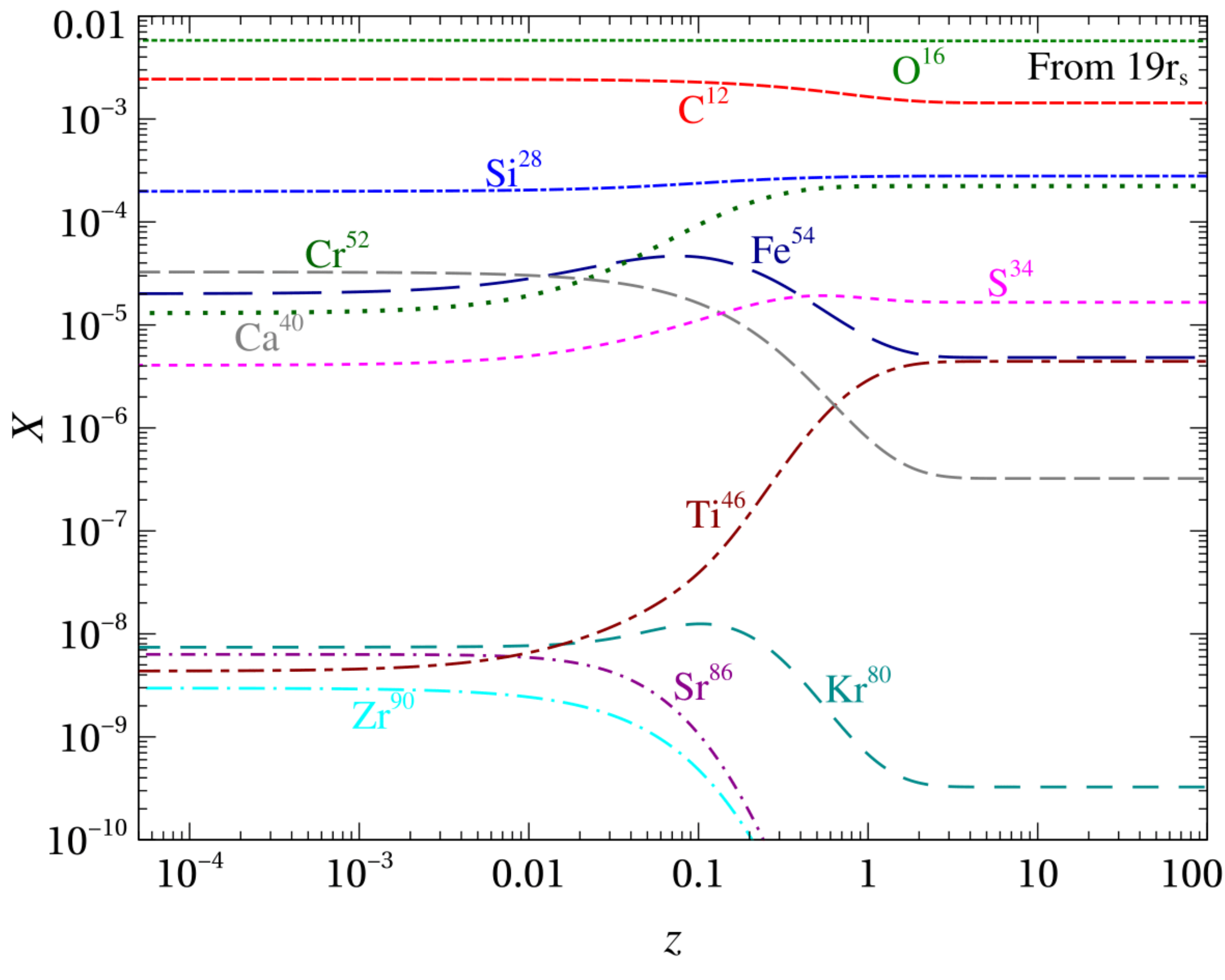}
    \caption{Same as Fig. \ref{outflow set1 rea 0.1 17 rs}, except the launching radius 19$r_s$.}

\label{outflow set1 rea 0.1 19 rs}
\end{figure}

\begin{figure}

	\includegraphics[width=\columnwidth]{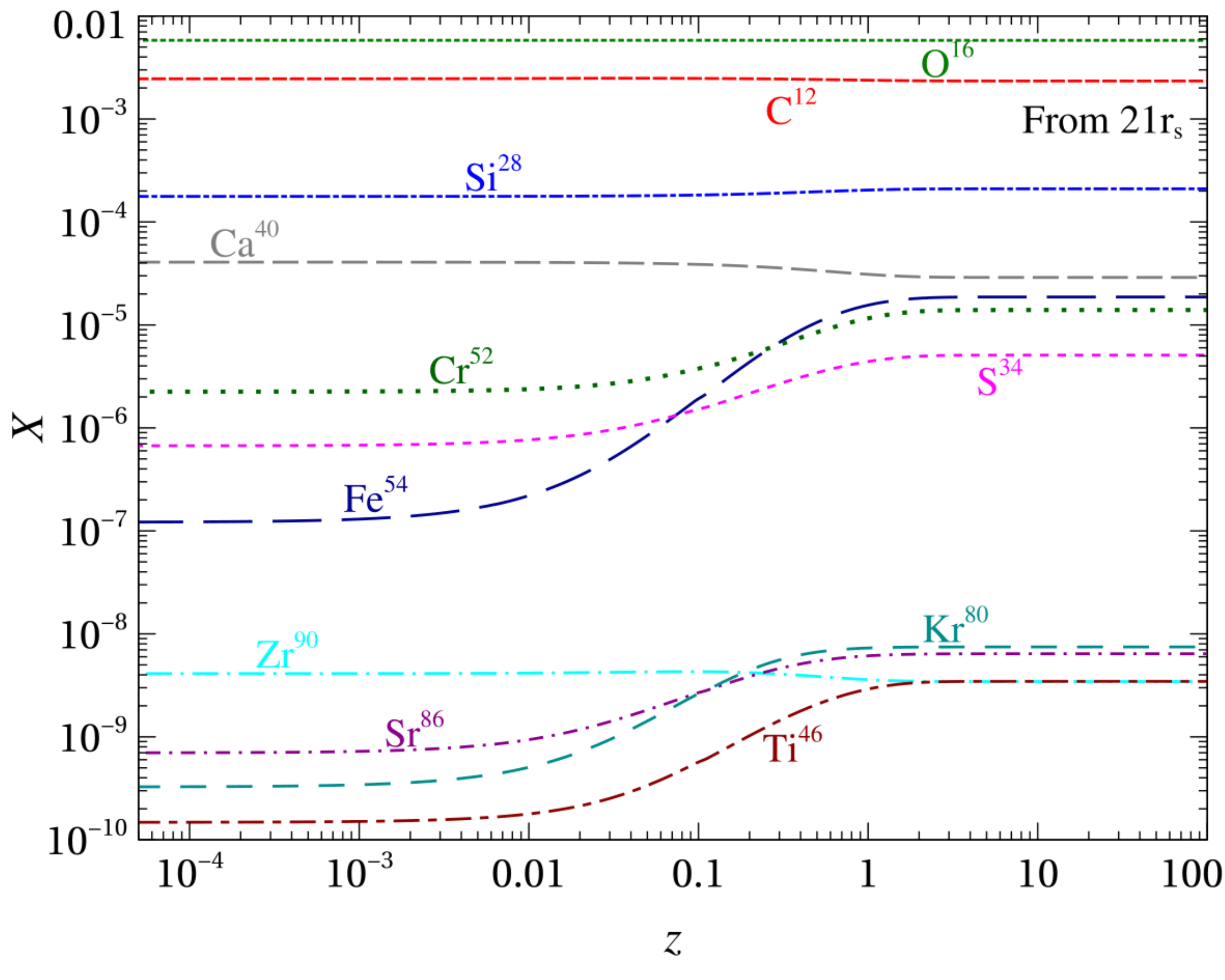}
    \caption{Same as Fig. \ref{outflow set1 rea 0.1 17 rs}, except the launching radius 21$r_s$.}

\label{outflow set1 rea 0.1 21 rs}
\end{figure}

However, when the outflow launches from the middle region of the disc, significant nucleosynthesis takes place and abundances evolve drastically. These abundance evolutions are shown in Figs. \ref{outflow set1 rea 0.1 17 rs}, \ref{outflow set1 rea 0.1 19 rs} and \ref{outflow set1 rea 0.1 21 rs}. These are some typical abundance evolutions, when respective outflows are launched from different radii of the middle region of Set 1. However, the respective choices of launching radii are arbitrary, when for the present purpose the goal is not understanding the outflow mechanism. As mentioned above, initial abundances for nucleosynthesis in outflow are fixed by the respective disc's abundances at the launching radius. As in the middle region of the disc itself, abundances evolve very fast, initial abundances for outflow from different radii become very different. This makes large difference in abundance evolution of different elements in outflows launched from different radii of the middle region. As we see that for the outflow from 17$r_s$ (Fig. \ref{outflow set1 rea 0.1 17 rs}), except $Si^{28}$, abundances of all the elements decrease. For outflow from 19$r_s$ (Fig. \ref{outflow set1 rea 0.1 19 rs}) abundances of $S^{34}$, $Ca^{40}$ and $Ti^{46}$ increase. However, for outflow from 21$r_s$ (Fig. \ref{outflow set1 rea 0.1 21 rs}), most of the elements shown are not only survived from the disc but also synthesized in the outflow. Nevertheless, we find that whenever outflow is launched from the middle region, the final abundances of Mg, Si, Ar, Ti and Cr remain higher than the solar abundances significantly in most of the cases. For the Set 3 disc, the same scenario repeats. \\

  \subsection{AGN}
  
\begin{figure}

	\includegraphics[width=\columnwidth]{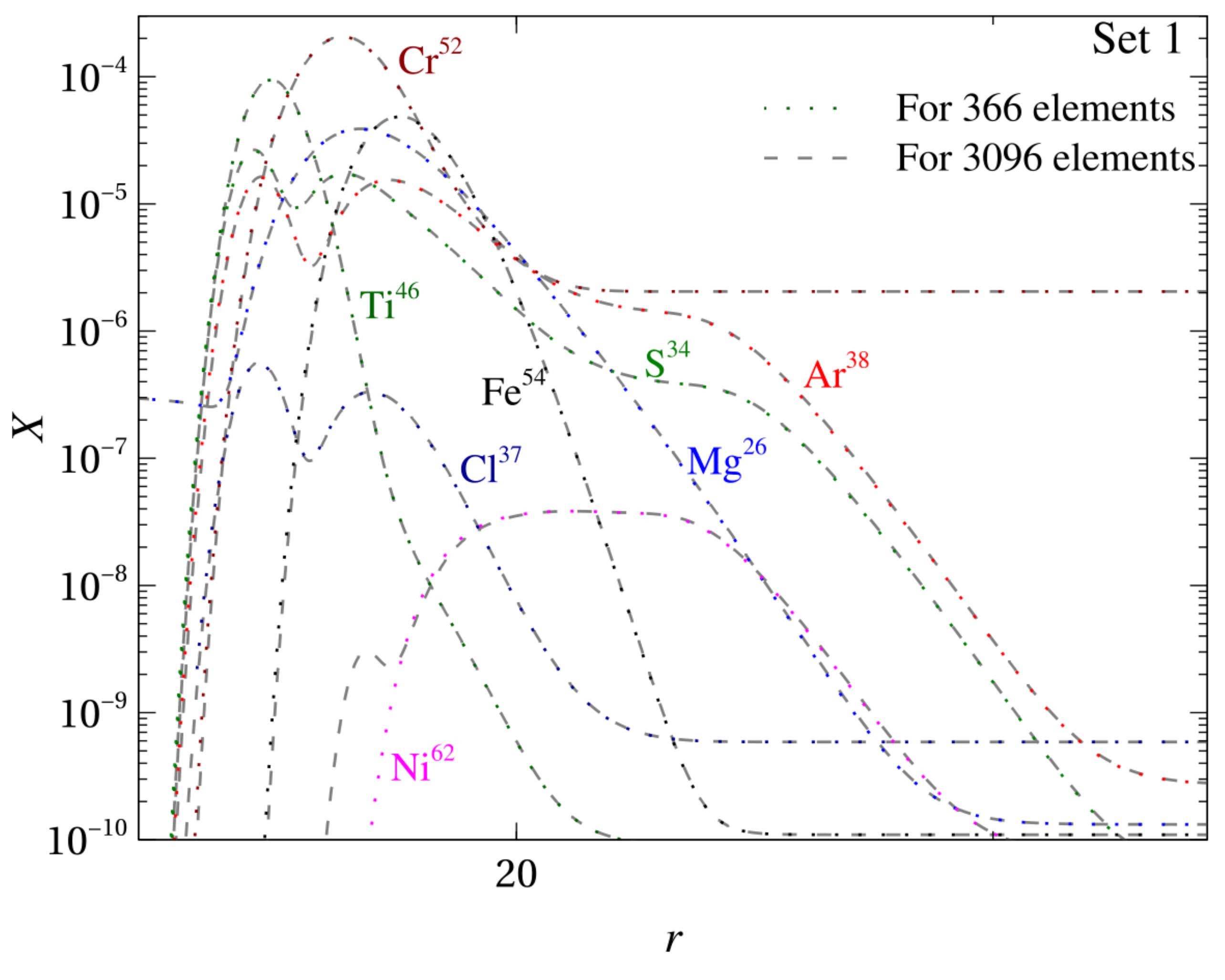}
    \caption{Comparison of abundance evolutions of some elements in XRB for Set 1 between 366 and 3096 nuclear networks.}

\label{comparison 366 3096}
\end{figure}

To simulate the nucleosynthesis of advective discs in AGN in reasonable time, we consider the nuclear network of 366 elements which includes up to different isotopes of Yttrium. The choice is justified as we notice that elements with higher mass number are not produced significantly in XRBs. Indeed, we compare nucleosynthesis in XRBs with 366 nuclear network with that of 3096 network, shown in Fig. \ref{comparison 366 3096}. It shows that all the abundances match quite well for two nuclear networks, except that we lose some information of heavier elements which have at the most $X \sim 10^{-8}$.
  
  		\subsubsection{Nucleosynthesis in disc}
  
\begin{figure}

	\includegraphics[width=\columnwidth]{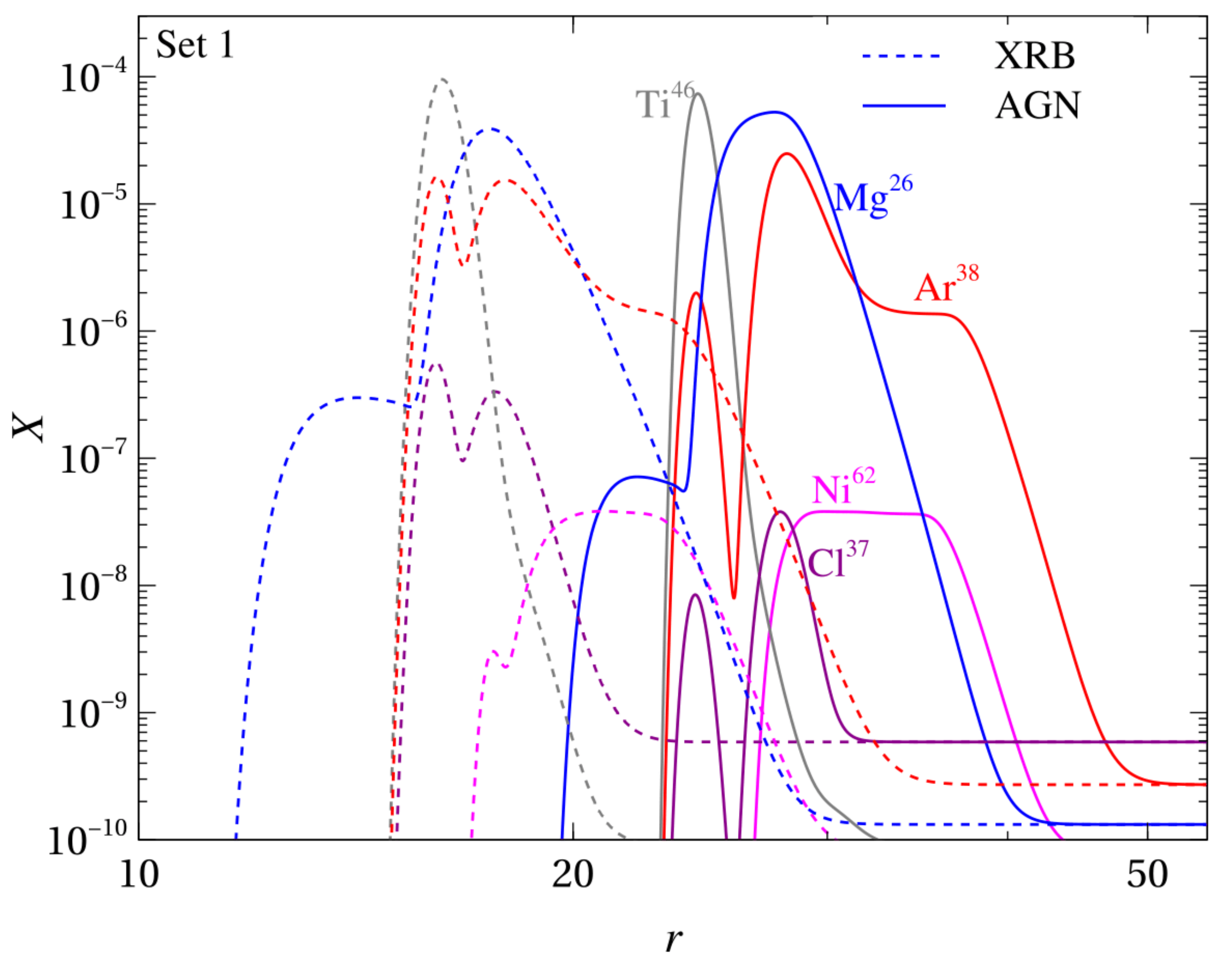}
    \caption{Comparison of abundance evolutions of some elements in Set 1 between XRB and AGN.}

\label{comparison bi AGN}
\end{figure}

The abundance evolution of different elements looks similar but with some differences in the evolutionary pattern compared to XRBs. Also interestingly, the location of middle region where most of the nucleosynthesis taking place shifts outward, as shown in Fig. \ref{comparison bi AGN}. This is because the residence time of matter increases for AGNs, hence all the disintegration and evolution of abundances occur at larger radii than XRBs. Rest remains same.

         \subsubsection{Nucleosynthesis in outflow}
         
\begin{figure}

	\includegraphics[width=\columnwidth]{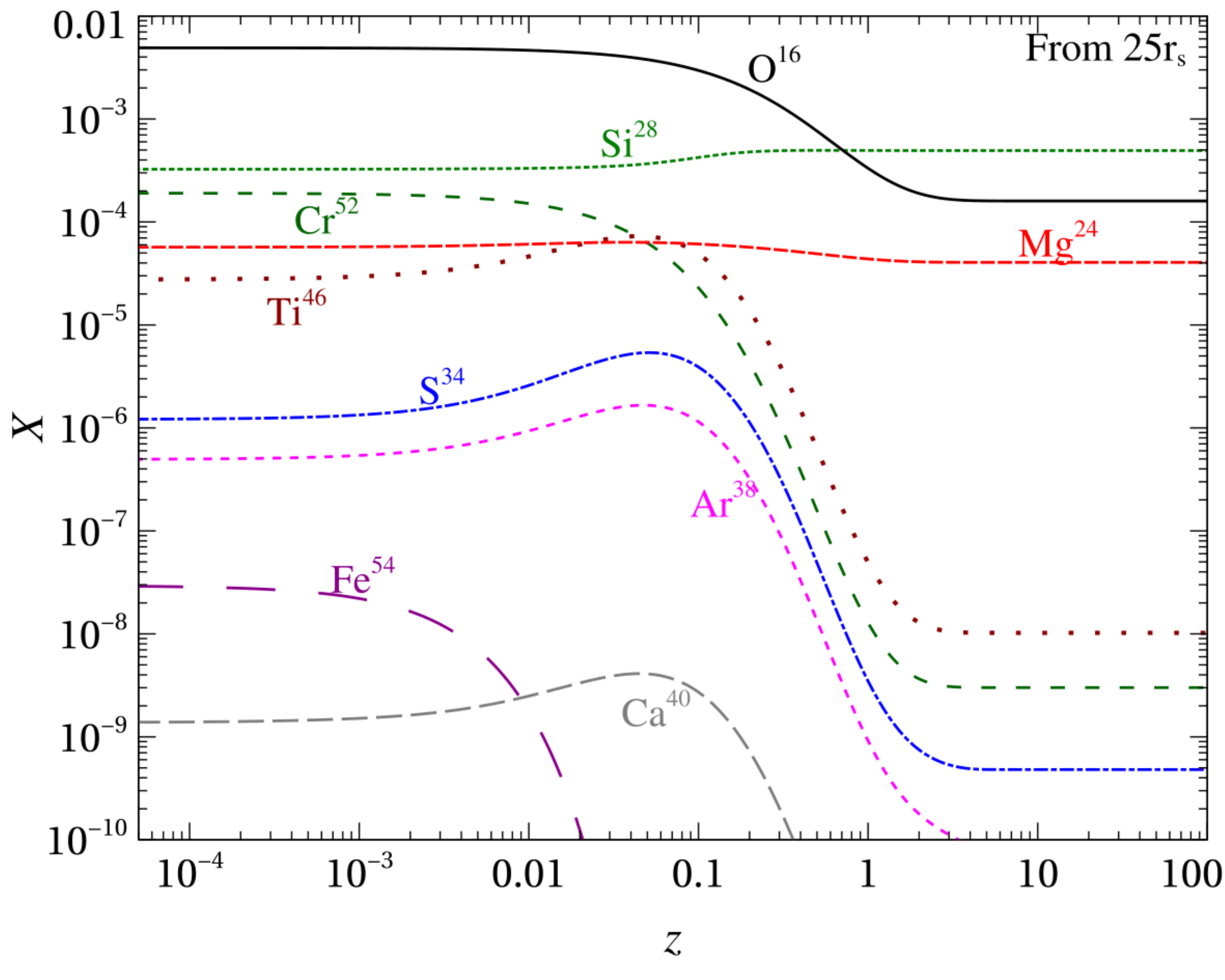}
    \caption{Evolution of abundances when outflow is launched from 25$r_s$ of Set 1 in AGN with $\dot{M}_{ej}/\dot{M}_{acc}=0.1$. Here $z$ is the distance from accretion disc.}

\label{outflow AGN set1 rea 0.1 25 rs}
\end{figure}

\begin{figure}

	\includegraphics[width=\columnwidth]{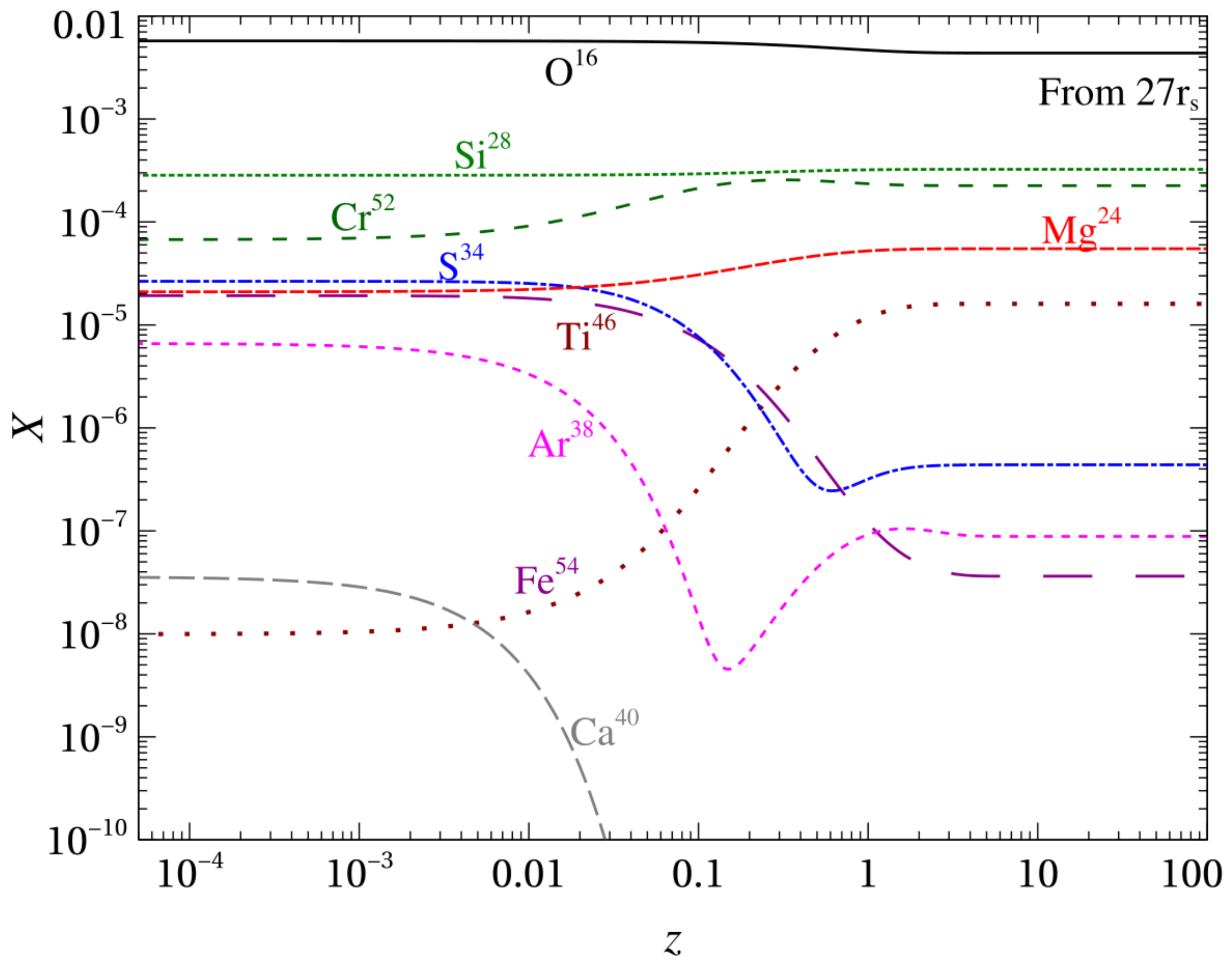}
    \caption{Same as Fig. \ref{outflow AGN set1 rea 0.1 25 rs}, except the launching radius 27$r_s$.}

\label{outflow AGN set1 rea 0.1 27 rs}
\end{figure}

\begin{figure}

	\includegraphics[width=\columnwidth]{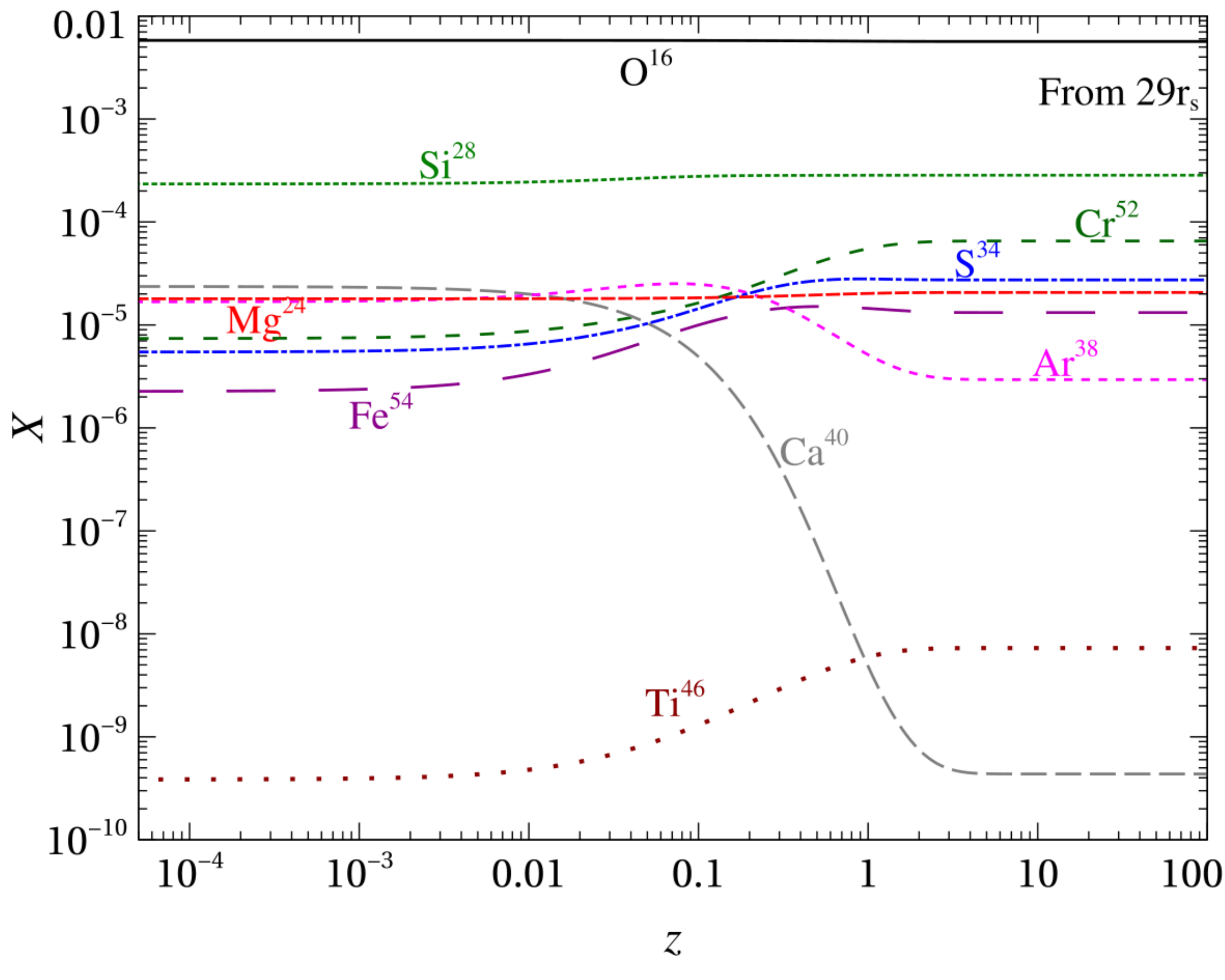}
    \caption{Same as Fig. \ref{outflow AGN set1 rea 0.1 25 rs}, except the launching radius 29$r_s$.}

\label{outflow AGN set1 rea 0.1 29 rs}
\end{figure}

Like XRBs, nucleosynthesis mostly occurs in the outflow, when outflow is launched from the middle region of the advective disc. The abundance evolutions in outflow, when it is launched from different radii of Set 1, are shown in Figs. \ref{outflow AGN set1 rea 0.1 25 rs}, \ref{outflow AGN set1 rea 0.1 27 rs} and \ref{outflow AGN set1 rea 0.1 29 rs}. Here in most of the cases, final abundances of Mg, Si, Ti, Cr remain higher than solar abundance.\\

\begin{figure}

	\includegraphics[width=\columnwidth]{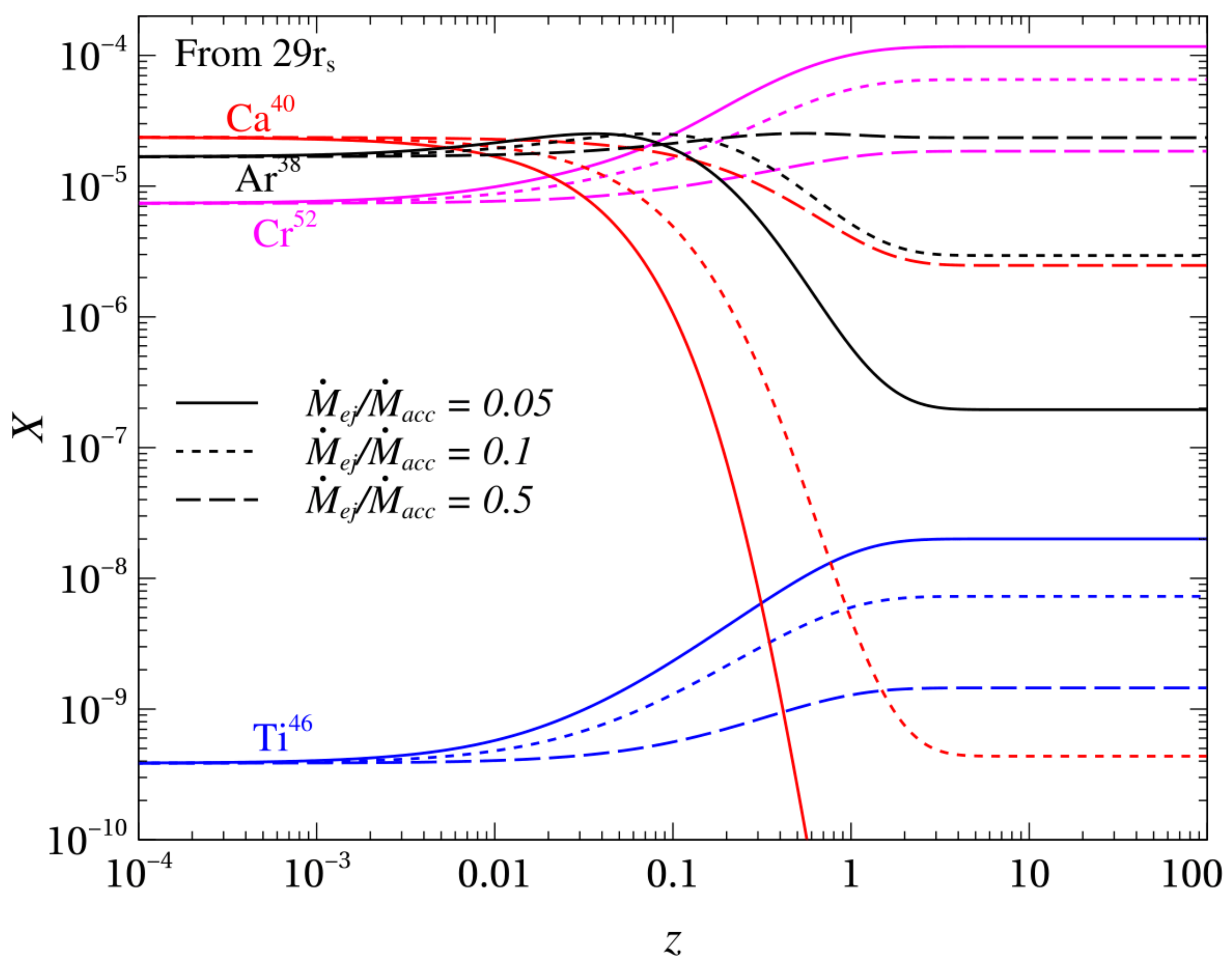}
    \caption{Evolution of some representative abundances when outflow is launched from $29r_s$ of Set 1 in AGN, with different $\dot{M}_{ej}/\dot{M}_{acc}=0.05$, 0.1, 0.5. Here $z$ is the distance from accretion disc.}

\label{different rea outflow AGN}
\end{figure}

One crucial point to note is that all the nucleosynthesis in outflow depends on the outflow velocity which is directly proportional to $\dot{M}_{ej}/\dot{M}_{acc}$. Observationally, the velocity of the outflow from advective disc is also not well constrained. Hence, we choose a typical value of $\dot{M}_{ej}/\dot{M}_{acc}=0.1$, i.e mass ejection rate is 10\% of mass accretion rate, in most of the computations which seems quite reasonable. However, we also compare nucleosynthesis in outflow with different values of $\dot{M}_{ej}/\dot{M}_{acc}$, shown in Fig. \ref{different rea outflow AGN}. With the increase of $\dot{M}_{ej}/\dot{M}_{acc}$ (hence, velocity of the outflow), nuclei get lesser time to interact and final abundance follows the initial launching abundance. It is clear from Fig. \ref{different rea outflow AGN} that larger the $\dot{M}_{ej}/\dot{M}_{acc}$, larger the velocity, lesser the change in abundance. This confirms that if the outflow velocity is large, final abundance will remain same as disc abundance. However, for lower velocity of outflow, final abundance can change drastically from the disc abundance.

\section{Caveats, Observational constraint and support}

\begin{table*}
	\centering
	\caption{Comparison of solar abundances of some key elements with abundances in the disc and outflow when it is launched from different radii.}
	\label{key elements}
	\begin{tabular}{cccccccc}
		\hline
		\multirow{2}{*}{Element} & \multirow{2}{*}{Solar abundance} & \multicolumn{2}{c}{$17r_s$} & \multicolumn{2}{c}{$19r_s$} & \multicolumn{2}{c}{$21r_s$}\\
		\cline{3-4} \cline{5-6} \cline{7-8}
		 & & Disc & Outflow & Disc & Outflow & Disc & Outflow\\
		 \hline\\
		Mg & $1.789\times10^{-5}$ & $6.946\times10^{-5}$ & $6.159\times10^{-5}$ & $3.190\times10^{-5}$ & $7.044\times10^{-5}$ & $1.908\times10^{-5}$ & $3.762\times10^{-5}$ \\
		Si & $1.749\times10^{-4}$ & $3.173\times10^{-4}$ & $4.955\times10^{-4}$ & $2.115\times10^{-4}$ & $3.208\times10^{-4}$ & $1.819\times10^{-4}$ & $2.249\times10^{-4}$ \\
		S & $1.359\times10^{-4}$ & $1.888\times10^{-5}$ & $2.801\times10^{-8}$ & $1.055\times10^{-4}$ & $1.730\times10^{-5}$ & $1.337\times10^{-4}$ & $9.192\times10^{-5}$ \\
		Ar & $6.913\times10^{-6}$ & $4.726\times10^{-6}$ & $7.562\times10^{-9}$ & $1.177\times10^{-5}$ & $3.194\times10^{-6}$ & $8.537\times10^{-6}$ & $1.402\times10^{-5}$ \\
		Ca & $4.144\times10^{-5}$ & $2.412\times10^{-6}$ & $8.101\times10^{-9}$ & $3.269\times10^{-5}$ & $5.282\times10^{-7}$ & $4.076\times10^{-5}$ & $2.898\times10^{-5}$ \\
		Ti & $1.586\times10^{-7}$ & $5.447\times10^{-5}$ & $7.367\times10^{-8}$ & $3.042\times10^{-7}$ & $3.160\times10^{-5}$ & $1.698\times10^{-7}$ & $3.295\times10^{-7}$\\
		Cr & $2.046\times10^{-6}$ & $2.338\times10^{-4}$ & $4.750\times10^{-10}$ & $1.337\times10^{-5}$ & $2.510\times10^{-4}$ & $2.453\times10^{-6}$ & $1.422\times10^{-5}$ \\
		Mn & $1.095\times10^{-5}$ & $2.238\times10^{-7}$ & $4.044\times10^{-17}$ & $5.064\times10^{-6}$ & $5.166\times10^{-8}$ & $1.055\times10^{-5}$ & $3.778\times10^{-6}$ \\
		Fe & $3.155\times10^{-4}$ & $1.239\times10^{-5}$ & $1.603\times10^{-22}$ & $3.129\times10^{-4}$ & $2.440\times10^{-5}$ & $3.164\times10^{-4}$ & $3.136\times10^{-4}$\\
		Co & $4.267\times10^{-6}$ & $2.528\times10^{-10}$ & $1.937\times10^{-30}$ & $1.753\times10^{-7}$ & $9.374\times10^{-11}$ & $3.406\times10^{-6}$ & $3.834\times10^{-8}$ \\
		\hline
	\end{tabular}
	
\end{table*}

There are state-of-art GRMHD simulations of black hole accretion including advection dominated accretion flow (ADAF) and radiative transfer (e.g. \citealt{2007MNRAS.375..513M,2012MNRAS.426.3241N,2014MNRAS.439..503S}),
which showed the time evolution and formation of quasisteady disc. However,
till now, to the best of our knowledge, no work on nucleosynthesis is
performed selfconsistently along with hydrodynamic simulation. Recent
venture of nucleosynthesis were mostly by performing full nuclear reaction
network on tracer particles in a post processing step after GRMHD simulation
(\citealt{2017PhRvL.119w1102S}). Needless to mention, following similar procedure for the 
present purpose will reveal more accurate results. However, throughout the whole time span
of a flow and nucleosynthesis, the most contributory
elemental abundances come from the quasi-steady part, because it 
remains for most of the time period. That is why, here we emphasize only
steady disc hydrodynamics profiles, which are indeed quite simpler to tackle without
loss of important physics and handy.

Through spectroscopic lines, winds are predicted observationally only in softer states of XRBs, when jet is quenched (\citealt {2012MNRAS.422L..11P}). In most of the cases, absorption line of Iron appears when an XRB changes its state from low/hard state (when advective disc dominates) to softer state (when Keplerian part of disc starts to dominate). There are many proposals to explain possible reasons for this change. Also the detection of a weak wind in the hard state of GRS 1915+105 (\citealt {2002ApJ...567.1102L}) indicates the possibility of ejection of significant amount of matter in both the soft and hard states. However, due to higher ionization in the hard state, the ions become fully stripped and wind becomes transparent and hence invisible (\citealt {2013AcPol..53..659D}). Nevertheless, \cite {2012ApJ...750...27N}, based on transition from hard to soft states in GX-13+1, showed that overionization can not be the only reason for invisibility of wind in a hard state. Another possible reason being thermodynamical instability of wind for the range of proper ionization parameter, which is suitable for the appearance of absorption lines in a hard state (\citealt {2013MNRAS.436..560C}). \cite {2012MNRAS.422L..11P} showed the dependence of angular inclination of the disc with the appearance of absorption line. Nevertheless, the theoretical modeling is in its infancy stage to conclude something about the invisibility of line as well as wind in a hard state. This is a crucial constraint for our work to say anything about the velocity of the outflow as well as final abundance in the outflow. However, we know from theoretical calculation that hot advective disc is more prone to outflow than cold Keplerian disc due to the presence of various additional physics in the
former (\citealt{1994ApJ...428L..13N, 1996ApJ...464..664C, 2003S&W....42j..98F, 2012MNRAS.426.3241N}; MM18).

Although the abundance evolution appears to be insignificant in order to observe it in the galactic scale, when we observe a particular LMXB, the evolution can change the observational analysis very much. From observational analysis and modeling, till now the understanding is that thermally driven wind is launched from $10^3-10^4r_s$ of a Keplerian disc (\citealt{2005A&A...436..195B,2012A&A...543A..50D,2013AcPol..53..659D,2018MNRAS.473..838D}). In most cases, highly ionized Iron lines only are detected. For very few sources, many spectroscopic lines are detected (\citealt {2004ApJ...609..325U, 2009ApJ...701..865K, 2012A&A...543A..50D}), and GX 13+1 is one of such sources. \cite {2004ApJ...609..325U} and \cite {2012A&A...543A..50D} reported larger [S/Fe] ratio than solar value repeatedly. Although \cite {2018ApJ...861...26A} attempted to model the disc wind with multiple absorption zones with fewer variable abundances, they acknowledged that complex modeling of absorption lines using single absorber with super-solar Ca, Ar, S, Si and Mg abundances is also a possible solution. Other source for which many spectroscopic lines are detected is GRO J1655-40 (\citealt {2009ApJ...701..865K}). The authors there reported that different models predicted overabundances of Cr, Mn and Co, relative to Fe, by at least 50\%. They tried to explain the abundance distribution as an effect of early supernova, though there are many discrepancies remain in the abundances of Ti, V and Co. It is quite unlikely that these overabundances of different elements, present in the wind, are also present in the companion star (\citealt {2015A&A...582A..81J}). Also we know that due to lower temperature, nucleosynthesis is not possible in a Keplerian disc. Therefore, the question remains, from where these overabundances of different elements in comparison with solar are arising. We show that these overabundances possibly arise due to nucleosynthesis in the advective part of disc and corresponding outflow, and enriching the atmosphere of outer Keplerian part of the disc by evolved elements, from where the wind is launched. From the position of underlying LMXB in the hardness-intensity diagram (\citealt {2012MNRAS.422L..11P}), we can infer that the disc is not in a complete soft state when we observe wind. It also contains significant advective component, however when jet is quenched. Moreover, observationally, it is not possible to detect isotopic lines. Therefore, we compute total isotopic abundances of different elements in the disc-wind system. Abundances of different key elements in the disc and outflow for $\dot{M}_{ej}/\dot{M}_{acc}=0.1$, when launched from different radii, along with respective solar values are listed in Table \ref{key elements}. We find that abundance of Fe becomes highly sub-solar when outflow is launched from the inner portion of the middle region. The possible overabundances of Mg, Si, Ar and Cr can be easily explained by the present work. The range of observationally predicted overabundance is also $2-6$ times solar abundance, which matches with our results. There is a predicted overabundance of Ti also from our calculation, which is not detected yet. Again observationally, the predicted overabundances of S, Ca, Mn and Co in some cases are possibly implying the complex dynamics of disc, outflow and wind which we are unable to capture yet. However, as Iron abundance decreases, larger [S/Fe] ratio becomes possible for certain cases. We also find that in the inner region of the advective disc, due to higher temperature, mostly free nucleons are there. It may also be one possible reason for not to see spectroscopic lines in a pure hard state.

In most of the sources (\citealt {2012MNRAS.422L..11P}) only Fe XXV and Fe XXVI ionized spectroscopic lines are present. Then to find out the column density and other physical variables of wind as well as disc, metallicity comes in the picture and till now observers use solar metallicity. Actually, we need to know how much fraction Iron is of total. If by the nucleosynthesis in advective disc and outflow, abundance of Iron changes, which is the case in the inner region of disc, then all the analysis will change drastically. 

For AGN, the disc and outflow structures are more complex and involved. There are many processes, specially galactic chemical evolution (\citealt {1999ARA&A..37..487H}), which can account for different elemental abundances for AGN. Therefore, the job of comparing quantitatively our nucleosynthesized abundances with observational signature is quite tedious and even nucleosynthesis effect may be insignificant in comparison with effects from other sources. Only we can conclude that this abundance change can be important when we try to model cosmological abundances of galaxies at high redshifts (\citealt{1989ApJ...336..572J}).
 
\section{Justification of outflow model and nucleosynthesis therein}
\subsection{Outflow model}
For the present purpose, we use a very simplistic model of outflow which assumes adiabaticity, spherical expansion and constant velocity. Fixing $\dot{M}_{ej}/\dot{M}_{acc}=0.1$ gives outflow velocity $\sim10^3$ Km/s, which is the typical wind velocity observed in XRBs (\citealt {2013AcPol..53..659D}). However, some estimates also argued that the mass outflow rate to be of the order of the mass accretion rate (e.g. \citealt {2004ApJ...609..325U, 2009ApJ...695..888U, 2012MNRAS.422L..11P}). Hence, the outflow velocity may also increase and the final abundance in the outflow may remain almost same as the disc abundance, as we discussed earlier. Nevertheless, GRMHD simulation of magnetized advective accretion disc 
(\citealt{2012MNRAS.426.3241N}) showed outflow velocity to be varying in the range $0.001-0.04c$ for the 
range of disc radius $50-5r_s$, depending on different configurations of simulation. 
Moreover, the authors there found that $\dot{M}_{ej}/\dot{M}_{acc}=0.1$ at $r=20r_s$. Therefore, our chosen 
outflow velocities are quite consistent with the simulation, which confirms the robustness of final 
nucleosynthesized products.

\subsection{Sensitivity to temperature}
Nucleosynthesis is very much sensitive to temperature and our all the 
evolutions of abundances take place in a temperature range $(3-6) \times 10^9 K$. 
Therefore, wherever this temperature lies in the disc, the evolution of elements 
occurs with the change in abundances in the disc. However, as mentioned above, GRMHD simulation of 
magnetized advective accretion disc (\citealt{2012MNRAS.426.3241N}) showed that outflow could 
take place from a large range of radius of the whole disc. Interestingly, the temperature range, 
in which nucleosynthesis is most sensitive, lies between two extrema: virial temperature in a 
typical low/hard state and blackbody thermal temperature of the Keplerian disc in a high/soft state. 
Moreover, the positions of sources exhibiting wind in the Q-diagram 
(\citealt{2012MNRAS.422L..11P}) indicate the wind to be detected in between these two states. 
Obviously, if the outflow takes place from a radius where temperature is different from 
$(3-6) \times 10^9 K$, the nucleosynthesis in the outflow presented here is less relevant. 
However, the position of sources with wind in the Q-diagram and
super-solar abundances of different elements together suggest outflow to take place
from the said specific region of the disc. Nevertheless, performing nucleosynthesis
in post processing step after performing threedimensional GRMHD simulation of magnetized
advective accretion disc might give resolution to this confusion, which can
be explored in future.

\subsection{Comparison between nuclear and outflow timescales}
There are numerous reactions occurring in the outflow. One typical chain of reactions, which 
is crucial for the production of $Cr^{52}$ from $Fe^{54}$ is: $Fe^{54}(H^1)Mn^{53}(H^1)Cr^{52}$. Now, 
we take $Fe^{54}(H^1)Mn^{53}$ as a typical nuclear reaction in the outflow. 
The reaction rate of this reaction at $3.5\times 10^9$ K, which is the typical temperature at the 
base of outflow, is ${\cal R} \sim 10^3 s^{-1}$. Therefore, the typical nuclear time scale is 
$t_{nuc}=1/{\cal R}=0.001$ s. Now most of the nucleosynthesis in the outflow takes place within 
the distance of $2 r_s$ from the disc and a typical velocity of outflow is $0.0045c$. This gives the 
outflow timescale $\sim 0.044 s$, which is much larger than $t_{nuc}$. This indicates that 
significant nucleosynthesis could take place in the outflow itself. As the temperature decreases 
with distance from the disc, nuclear timescale increases and no significant nucleosynthesis could 
take place above $2 r_s$. The temperature at the base of the outflow is already constrained by 
the temperature of the disc. Therefore, for so short distance, exact variation of temperature does not 
affect largely the abundances, which supports our simplistic parameterized outflow model.

\section{Conclusions}

We have performed nucleosynthesis in the advective disc in XRBs and AGNs, and subsequently in the respective outflows launched from there rigorously. Starting from solar abundance, elements have evolved in the disc. The velocity of outflow is a very key parameter in order to determine final abundances and till now the outflow velocity is not constrained observationally. However, comparison with outflow velocity from GRMHD simulation indicates that chosen outflow velocity is quite reasonable. Inclusion of a larger nuclear network does not work to synthesize heavier elements in the accretion disc. The abundance evolution in the outflow has confirmed that it is very crucial to consider nucleosynthesis in the outflow when its velocity is low. All the earlier works neglected outflow nucleosynthesis. The present work has shown that neglecting outflow nucleosynthesis may severely lead to wrong conclusion about final abundances. In other words, inference about abundance evolution based on disc nucleosynthesis alone is misleading.\\ 

We have found that when outflow is launched from a suitable region of the disc with $\dot{M}_{ej}/\dot{M}_{acc}=0.1$, final abundances of the elements Mg, Si, Ar, Ti and Cr remain super-solar, irrespective of the exact position from where the outflow is launched for most of the cases. In the observation of wind from LMXBs, when many spectroscopic lines are available, people attempted to model wind with the assumption of either one absorption zone or many absorption zones. If we combine the latest observations, then the one zone model requires super-solar abundance of Mg, Si, S, Ar, Ca, Cr, Mn, Co. We have proposed that these overabundances arise due to the nucleosynthesis happening in the advective disc of XRBs and outflows from there. Required overabundances of Mg, Si, Ar and Cr and larger ratio of [S/Fe] have matched certainly that obtained from our simulation. Rest of the over-abundant elements indicate possibly complex dynamics of the disc and outflow, and also the subtle dependence on the velocity of outflow, but possibly uncaptured yet. Also we have taken the initial abundance of the accretion disc as solar abundance. It may vary from star to star depending on the companion. In future, we will study whether the abundance evolution in outflows changes or not, if the initial abundance for nucleosynthesis in disc changes largely depending on the different evolutionary stages of the companion.

As all the elements are disintegrated into free nucleons in the inner region of the disc, this can be the possible reason for invisibility of wind, i.e. spectroscopic lines, in pure hard states. Another possible reason is that the density of the advective disc is $5-6$ orders of magnitude less than Keplerian disc. When an outflow takes place from an advective disc, after advancing some distance it will be cooled down and nuclei will capture electrons, becoming eligible for producing spectroscopic lines. However, by that time, density becomes too low to produce any signature of absorption. Nevertheless, in many observations when only Iron lines are present, these nucleosynthesized abundances have the potential to change the whole observational analysis. Above all, finally we can say that this nucleosynthesis argues for the modeling of wind from accretion disc more complex and open.
%Normally the next section describes the techniques the authors used.
%It is frequently split into subsections, such as Section~\ref{sec:maths} below.

%\subsection{Maths}
%\label{sec:maths} % used for referring to this section from elsewhere

%Simple mathematics can be inserted into the flow of the text e.g. $2\times3=6$
%or $v=220$\,km\,s$^{-1}$, but more complicated expressions should be entered
%as a numbered equation:

%\begin{equation}
%    x=\frac{-b\pm\sqrt{b^2-4ac}}{2a}.
%	\label{eq:quadratic}
%\end{equation}

%Refer back to them as e.g. equation~(\ref{eq:quadratic}).

%\subsection{Figures and tables}

%Figures and tables should be placed at logical positions in the text. Don't
%worry about the exact layout, which will be handled by the publishers.

%Figures are referred to as e.g. Fig.~\ref{fig:example_figure}, and tables as
%e.g. Table~\ref{tab:example_table}.


%\begin{table}
%	\centering
%	\caption{This is an example table. Captions appear above each table.
%	Remember to define the quantities, symbols and units used.}
%	\label{tab:example_table}
%	\begin{tabular}{lccr} % four columns, alignment for each
	%	\hline
	%	A & B & C & D\\
	%	\hline
	%	1 & 2 & 3 & 4\\
	%	2 & 4 & 6 & 8\\
	%	3 & 5 & 7 & 9\\
%		\hline
%	\end{tabular}
%\end{table}

%The last numbered section should briefly summarise what has been done, and %describe
%the final conclusions which the authors draw from their work.

\section*{Acknowledgements}

SRD would like to thank Susmita Chakravorty and Gajendra Pandey for helpful discussions. Thanks are
also due to the anonymous referee for thorough reading the manuscript with suggestions
to improve the presentation. This work is partly supported by the fund of DST INSPIRE fellowship belonging to SRD and partly by the project
with research Grant No. DSTO/PPH/BMP/1946 (EMR/2017/001226).

%The Acknowledgements section is not numbered. Here you can thank helpful
%colleagues, acknowledge funding agencies, telescopes and facilities used etc.

%%%%%%%%%%%%%%%%%%%%%%%%%%%%%%%%%%%%%%%%%%%%%%%%%%



\bibliographystyle{mnras}
\bibliography{adv_nucleosynthesis} % if your bibtex file is called example.bib

%\begin{thebibliography}{99}
%\end{thebibliography}

%%%%%%%%%%%%%%%%%%%%%%%%%%%%%%%%%%%%%%%%%%%%%%%%%%

%%%%%%%%%%%%%%%%% APPENDICES %%%%%%%%%%%%%%%%%%%%%

\bsp	% typesetting comment
\label{lastpage}
\end{document}